\pdfoutput=1

\documentclass[12pt,preprint]{aastex}

\slugcomment{astro-ph/0701474v4 Nov 2}

\shorttitle{Interpretation of the Helix Planetary Nebula}
\shortauthors{Gibson $\&$ Schild}

\begin{document}


\title{Interpretation of the Helix Planetary Nebula using
Hydro-Gravitational-Dynamics: Planets and Dark Energy}


\author{Carl H. Gibson\altaffilmark{1}}
\affil{Departments of Mechanical and Aerospace Engineering and Scripps
Institution of Oceanography, University of California,
       San Diego, CA 92093-0411}

\email{cgibson@ucsd.edu}

\and

\author{Rudolph E. Schild}
\affil{Center for Astrophysics,
       60 Garden Street, Cambridge, MA 02138}
\email{rschild@cfa.harvard.edu}


\altaffiltext{1}{Center for Astrophysics and Space Sciences, UCSD}


\begin{abstract} 
Hubble Space Telescope (HST/ACS) images of the
Helix Planetary Nebula (NGC 7293) are interpreted using the
hydro-gravitational-dynamics theory (HGD) of Gibson 1996-2006.  HGD  
claims that
baryonic-dark-matter (BDM) dominates the halo masses of galaxies (Schild 1996)
as Jovian (Primordial-fog-particle [PFP])  Planets  (JPPs) in
proto-globular-star-cluster (PGC) clumps for all galaxy halo diameters bounded by stars. 
 From HGD, supernova Ia (SNe Ia) events always occur in planetary nebulae (PNe) within PGCs.  
The dying central star of a PNe slowly accretes JPP mass to grow the white-dwarf to $1.44 M_{\sun}$ 
instability from  $\ge 1000 M_{\sun}$ BDM within
luminous PNe diameters.  Plasma jets, winds and radiation driven by contraction and spin-up
of the carbon star evaporate JPPs revealing its Oort accretional cavity.
 SNe Ia events may thus be obscured or not obscured by radiation-inflated JPP atmospheres 
 producing systematic SNe Ia distance errors, so the otherwise mysterious  ``dark energy'' 
 concept is unnecessary.   HST/ACS and WFPC2  Helix images  show
$>7,000$ cometary globules and SST/IRAC images show $>20,000-40,000$, 
here interpreted as gas-dust cocoons of 
JPPs evaporated by the spin powered radiation of the PNe central
 white-dwarf.  Observed JPP masses 
$\approx 3 \times 10^{25}$ kg   with spacing
$\approx 10^{14}$ m for galaxy star forming regions give a 
density $\rho$ that fossilizes the primordial density  
$\rho_{0} \approx 3 \times 10^{-17}$ kg m$^{-3}$ existing for 
times $10^{12} \le t \le 10^{13} $ s  when the   plasma 
universe fragmented into  proto-superclusters, proto-clusters, and proto-galaxies.   
Pulsar scintillation spectra support the postulated multi-planet atmospheres. 
  
\end{abstract}




\section{Introduction}

Brightness values of Supernovae Ia (SNe Ia) events, 
 taken as standard candles for redshift
values $0.01 < z < 2$,  depart significantly from
those expected for the decelerating expansion rate of a flat universe  \citep{rie04}. 
Departures indicate a recent dimming at all frequencies by about
$30\%$, but with large scatter attributed to uncertainty in 
the SNe Ia models. For $\ge$15 years of study such evidence has accumulated with no
explanation other than an accelerating(!)
expansion rate of the universe, presumed to reflect a negative pressure from 
a time dependent $\Lambda (t)$ ``cosmological
constant'' component of the Einstein equations termed ``dark-energy'' \citep{bea05}.  
 Hubble Space
Telescope Advanced Camera for Surveys (HST/ACS) images have
such high signal to noise ratios  that both the scatter and the dimming
are statistically significant over the full range of $z$ values.  Bright SNe Ia observed for
$z \ge 0.46$ exclude ``uniform grey dust'' systematic errors, 
supporting a flat universe deceleration of expansion rate until the recent
``cosmic jerk'' to an accelerated expansion rate for $z
\le 0.46$.  This physically mysterious  ``dark energy''  interpretation
is made from SNe Ia observations
 because no alternative exists
in the commonly accepted (but fluid mechanically untenable) 
$\Lambda$-cold-dark-matter ($\Lambda \rm CDM$) hierarchically clustering
cosmological theory ($\Lambda \rm CDMHCC$, Table 2).    
In the following we suggest that the primordial exoplanets of HGD
 provide just the fluid-mechanically-correct alternative
demanded by the observed intermittent  SNe Ia brightness dimming.   New
physical laws are not required by HGD but $\Lambda \rm CDMHCC$ must be discarded.  The choice
is between planets and dark energy.  

When a massive primordial-planet (exoplanet, rogue-planet, planemo, Jovian) population comprised of 
 plasma-fossil-density ($\rho_0$ frozen H-He) PGC clumps of JPPs is recognized as the  
baryonic dark matter (BDM) and the interstellar medium (ISM), new scenarios
 are required for planetary nebulae (PNe) formation 
and for star formation, star evolution, and star death.  
From HGD the average galaxy BDM-ISM mass is $\approx 30 \times$  
the luminous  mass of stars and $\gg$  the more diffusive NBDM-ISM (CDM) mass.  
The million trillion-planet-PGCs per galaxy each have  fossil-density $\rho_0 \approx 
 \rho_{PGC} \approx 10^4$ 
greater than the $\langle \rho_g \rangle \approx 10^{-21}$ kg m$^{-3}$ galaxy  average and 
$\approx 10^9 \langle \rho_U \rangle$, where $\langle \rho_U \rangle \approx 10^{-26}$ kg m$^{-3}$ is the
flat universe average density at the present time.
 All stars and all JPP planets are thus born and grown within such $10^6 M_\sun$ 
 PGCs by mergers and accretions
  of PFP and JPP planets from the large PGC supply.  By gravity the planets 
 collect and recycle the dust and water of exploded stars to explain solar terrestrial planets and life.  
Application of fluid mechanics to the big bang, inflation, and the
plasma- and gas- self-gravitational structure formation epochs (Table 1) is termed 
hydo-gravitational-dynamics (HGD) theory (Gibson 1996, 2000, 2001, 2004, 2005).  Viscosity, density
and expansion-rate fix plasma gravitational fragmentation scales with a linear-spiral 
 weak-turbulence protogalaxy geometry \citep{gib06a,nom98}.   Voids between the
 $10^{46}-10^{43}$ kg   $\rho_{0}$ supercluster-to-protogalaxy 
 fragments first fill in the plasma epoch and later 
empty in the gas epoch by diffusion of a 
more massive neutrino-like population 
($\Omega_{NB} + \Omega_{B}=1, \Omega_{\Lambda} =0;\Omega=\langle \rho \rangle/\langle \rho_U \rangle$).
 Instead of  concordance cosmology values
($\Omega_{NB}=0.27,   \Omega_{B}=0.024, \Omega_{\Lambda} =0.73$)
commonly used \citep{wis07}, HGD
gives ($\Omega_{NB}=0.968,   \Omega_{B}=0.032, \Omega_{\Lambda} =0$).  Supercluster voids with 
radio-telescope-detected scales $\ge 10^{25}$ m at redshift $z \le 1$  
 \citep{rud07} confirm this prediction of
  HGD, but decisively contradict $\Lambda$CDMHC  where
  superclustervoids form last rather than first.

Many adjustments to standard cosmological, astrophysical, and astronomical
  models are required by HGD (Gibson 2006a, Gibson 2006b) and 
  many puzzling questions are answered. For example, why are
  most stars binaries and why are population masses of small stars larger than population masses
  of larger stars ($m_{SS} \ge m_{LS}$)?  From HGD it is because all stars form and grow
   by a frictional-clumping binary-cascade
  from small PFP planets.  Where do planetary nebulae and supernova remnants get their masses? 
  From HGD these masses are mostly evaporated ambient BDM planets, just as  the masses
  for many stars assumed larger than $ 2 M_\sun$ 
 can be attributed mostly to the brightness of the huge ( $\ge10^{13}$ m) 
 dark-matter-planet atmospheres they evaporate
 ($\S3$ Fig. 3).    
  Masses of supergiant OB and Wolf-Rayet stars are vastly overestimated 
  neglecting evaporated JPP brightness \citep{mau04,shi90}.   
  Because stars form from planets the first stars 
  must be small and early.   Large
  $\ge 2 M_\sun$ population III stars forming directly from $10^6 M_\sun$ Jeans mass gas clouds
   in CDM halos never happened and neither did re-ionization of the gas-epoch  
   back to a second plasma-epoch.  CDM halos never happened.  There were no dark ages because
   the first stars formed immediately from merging PFP planet-mass 
   clouds before the luminous hot gas cooled (at $\approx 10^{13}$ s).  
   All big stars were once little stars.  All little stars were once planets.  The
   mystery of massive low surface brightness galaxies \citep{one07} is solved.   From HGD
   these are protogalaxies where, despite maximum
    tidal agitation, the central PGCs have remained in their 
   original starless BDM state.  All proto-galaxies were created 
   simultaneously without stars at the end of the plasma epoch (at $\approx 3 \times 10^{5}$ y).

Jovian rogue planets dominating inner halo galaxy mass densities \citep{gib96} matches an identical,
but completely independent, interpretation offered from Q0957+561A,B 
quasar microlensing observations \citep{sch96}.  Repeated, continuous, redundant observations
 of the Q0957 lensed quasar for $> 20$ years by several observers and telescopes
confirm that the mass of galaxies
within all radii containing the stars must be dominated by 
planets \citep{ColS03,sch04a,sch04b,gib06a,gib06b}.  The non-baryonic
dark matter (NBDM) is probably a mix of neutrino flavors, mostly primordial and 
sterile (weakly collisional) 
with mass $m_{NBDM} \approx 30 m_{BDM}$.  NBDM is
super-diffusive and presently forms large outer galaxy halos and galaxy 
cluster halos.  From HGD, the function of NBDM is to continue the decelerating expansion
of the universe toward zero velocity (or slightly less) by  large scale gravitational forces.  A matter
dominated $\Lambda =0$ flat expanding universe monotonically decelerates from general relativity theory.  The entropy produced by its big bang turbulent beginning  \citep{gib05} implies a closed contracting fate for the universe, not an open accelerating expansion driven by  dark energy \citep{bus07}.

According to HGD, SNe Ia explosions always occur in PNes within PGC 
 massive dense clumps of  frozen primordial
planets where virtually all stars  form and die.  Planetary nebulae are not
just brief puffs of illuminated gas and dust ejected from dying stars in a vacuum, but are manifestations of
 $\approx 3 \times 10^{7}$  primordial dark matter planets per star in  galaxies of 
$3\%$ bright and $97\%$ dark PGCs.   
 When  the  JPP supply-rate of H-He gas is too small, stars die and
  cool as small helium or carbon white dwarfs. 
  With modest internal stratified turbulent mixing
  rates from larger JPP rates,
gravity compresses the carbon core, the angular momentum, and the magnetic field 
giving strong axial plasma jets and equatorial stratified turbulent plasma winds  \citep{gib07}.   
 Nearby JPP planets heated by the star and illuminated by the jets and winds
evaporate and become visible as a PNe with a white-dwarf central star.
PNes thus appear out of the dark whenever white-dwarf (WD) carbon stars 
are gradually growing
 to the Chandrasekhar limit of $1.44 M_{\sun}$ \citep{ham03,pen04,hac01}.  Gradual star
growth is dangerous to the star because its carbon core  may collapse from inadequate radial mass
mixing, giving a SNe Ia event. Modest star growth may be even more dangerous because
enhanced turbulence may mix and burn the carbon core but not mix the resulting
incombustible iron core,
which explodes at $1.4 M_{\sun}$ to form a neutron star in a supernova II event.  Rapid
stably stratified turbulent mixing within a star forced by
 a strong JPP rain may mix away both carbon core and iron core instabilities
to form $\gg 2 M_{\sun}$ superstars  \citep{ket06}.

Thus when JPPs in a PGC are 
agitated to high speed $V_{JPP}$,  the rapid growth of its stars
 will inhibit formation of  collapsing carbon cores 
  so that fewer carbon white dwarfs and fewer SNe Ia result.  More stars  
with $M_{Fe[Crit]} \approx  1.4 M_{\sun} <  1.44 M_{\sun}$ will centrally mix and explode 
as supernova II (SNe II) due to iron core collapse
to form neutron stars manifested as $1.4 M_\sun$ pulsars \citep{tho99}. Supernova II remnants
such as the Crab are mostly evaporated or evaporating JPPs.  
The speed $V_{JPP}$ of planets and planet clumps 
within a PGC, their spin, and the
size and composition of their gas-dust atmospheres are
critical parameters to the formation of larger planets, stars, and PNe, and will be
the subject of future studies (see Fig. 2 below).  These parameters are 
analogous to the small protein chemicals used for bacterial quorum sensing
in symbiotic gene expressions  \citep{loh03} by providing a form of PGC corporate memory.
Numerous dense, cold, water-maser and molecular-gas-clumps detected by radio telescopes
 in red giants and PNe
\citep{mir01,taf07} are massive ($\ge M_{Jup}$) JPPs and should be studied as such
  to reveal important JPP parameters such as $V_{JPP}$.  
 
A gentle rain of JPP comets permits the possibility that
a WD may grow its compressing carbon core to deflagration-detonation at the 
Chandrasekhar limit \citep{rop07}.  As the core
 mass and density grow, the angular momentum
  increases along with the strength of plasma beams
  and winds, giving strong increases in the 
 WD surface temperature, photon radiation, and observable PNe mass $M_{PNe}$.  Strong
 JPP rains lead to Wolf-Rayet (C, N and O class) stars
  cloaked in massive envelopes of evaporating
 JPPs misinterpreted as superwind ejecta.  With larger accretion rates, radially
 beamed internal wave mixing driven by buoyancy damped turbulence  \citep{kee05,gib06c, gib06d} 
 prevents both SNe II  detonation at $1.4 M_{\sun}$ and  
 SNe Ia detonation at $1.44 M_{\sun}$.  
 Turbulence and internal waves 
 cascade from small scales to large, mix and diffuse.  With strong forcing, turbulent combustion can be quenched  \citep{pet00}.  With moderate JPP accretion rates, white-dwarfs can  gradually burn
 gas to precisely enough carbon to collapse, spin up, and explode within the PNes they produce.
 
 It is known \citep{pad05} that Pre-Main-Sequence (PMS) star formation is
   not understood.  Numerical
  simulations from gas clouds and the Bondi-Hoyle-Littleton model of wake gas accretion
  show the larger the star the larger the gas accretion rate.  By conventional
  models of stars collapsing from clouds of molecular gas without planets
  there should be more large (supersolar) stars than 
  small (brown dwarf BD, red dwarf RD) stars, contrary to  
  observations \citep{cal05,alc00, gah07} that
    globulette-star-population-mass $m_{GS}$ decreases as 
    globulette-star-mass $M_{GS}$ increases; that is,
   $m_{G} \ge m_{BD} \ge m_{RD}  \ge m_{\sun}$. 
  Such a monotonic decrease in $m_{GS}$ with $M_{GS}$ giving  
  $m_{PFP} \ge m_{JPP}  \ge m_{BD}  \ge m_{RD}$ 
  is expected from HGD where all stars and planets form by hierarchical accretion from
   $10^{-3} M_{Jup}$ PFP dark matter planets within $3 \times 10^{15}$ m
   Oort cavities in the ISM.  Because $m=NM$, the decrease in $m_{GS}$ with $M_{GS}$
    is also supported by observations of exoplanet
   numbers $N$
   showing $dN/dM \sim M^{-1.2}$ for $M={(0.04-15)}M_{Jup}$ \citep{but06}, contrary
   to conventional planet formation models \citep{ida04,  bos01} where stars form
  mostly giant   $\ge M_{Jup}$ planets within $\le10^{12}$ m (10 AU) of the protostar.   
  
  Aging WR-stars, neutron stars, pulsars, and C-stars are most frequently
 identified in spiral-galaxy-disks (SGD) where tidal agitation 
 is maximum for the $\approx 10^{18}$ planets of galaxy BDM
  halos.  From HGD,  SGDs reflect PGC accretion from the halo.  As the initially 
  gaseous PGCs of protogalaxies freeze they become
  less sticky and collisional, so they diffuse out of their protogalaxy cores
 (with Nomura scale $L_N \approx 10^{20}$ m $\approx$ 2 kpc)    in growing orbits to 
 form the present   $\ge 30 L_N$ baryonic dark matter halos \citep{gib06a}.  
 Some tidally agitated PGCs become luminous as they are 
 captured and accreted back toward the original core region, forming SGD accretion  disks.  
 Dark or nearly dark PGCs leave thin great circle metal-free star wakes
 about the Galaxy center to $ (2-3) L_N$ radii, triggered and stretched away 
  by tidal forces
 \citep{gri06,bel06,ode01}, where the mass in these ancient star ``streams''  typically 
 exceeds that of their sources (eg.: the ``Orphan Stream'').   Dwarf galaxy clumps of PGCs in
 orbits out to $7 L_N$ leave streams of stars and globular clusters  \citep{iba02} and
 high-velocity-clouds of gas (HVCs) at the $15 L_N$
 distance of the Magellanic cloud stream \citep{iba07}.  Observations and HGD contradict
  suggestions of a non-baryonic dark matter origin or a capture origin for 
  these recently discovered Galactic objects.  Thousands of
  PGCs and their wakes covering $\approx 20 \%$ of the sky
   may be identified with anomalous velocity HVC objects and $\approx 200$
  isolated compact CHVCs covering $\approx 1 \%$ from their 
  neutral hydrogen signatures, masses $10^{4-5} M_{\sun}$,
  distances up to $6 \times 10^{21}$ m ($60 L_N$),
  and sizes $\approx 10^{17-18}$ m \citep{put02}.   Stars should form slowly enough to make
  white dwarfs with intermittently dimmed SNe Ia events in such gently agitated PGCs
   with small $V_{JPP}$ values.   Pulsars, however, always twinkle because lines
   of sight always intersect fossil electron density turbulence \citep{gib07} atmospheres 
   of planets powerfully evaporated by the SN II event and the pulsar  (see Fig. 12 $\S 4,\S 5$).
 
 From HGD, SNe Ia events will  be intermittently dimmed by Oort-rim-distant 
 JPP-atmospheres  evaporated  by the increasing radiation 
 prior to the event that has ionized and accreted all JPPs in the Oort cloud cavity. 
  Evidence
of a massive (several $M_{\sun}$) H-rich circumstellar medium at distances of up to nearly a 
light year ($10^{16}$ m) after the brightness maximum is indicated
by slow fading SNe Ia events \citep{woo04}.
 Our scenario of SNe Ia formation by gradual carbon WD growth of
  $\le M_{\sun}$ size stars
 fed by JPP comets contradicts the standard model for
  SNe Ia events (see $\S 2.3.1$) where  superwinds dump most of the mass 
  of  $(3-9) M_{\sun}$ intermediate size stars into the ISM.   
Few SNe Ia events are seen at large redshifts because billions
of years are needed to grow a $1.44 M_{\sun}$ star.  In  $\S 2.3.1$ we 
question PNe models involving intermediate size stars, their envelopes, and their superwinds.
Stars formed by a gassy merging planet clump binary 
cascade is supported by observations in star forming regions 
 of  ($10^{26} - 10^{29}$ kg) spherical
globulette objects with mass distributions dominated 
by the small mass globulettes \citep{gah07}.  

Theories describing the death of small to intermediate mass
 stars ($0.5 M_{\sun} - 9 M_{\sun}$) to form white-dwarfs and planetary nebulae
are notoriously unsatisfactory \citep{ib84}.   Neglecting the
ambient JPPs of HGD, observations of PNe mass and composition
indicate that most of the
matter  for such stars is inexplicably expelled \citep{kna82}  when they form dense carbon cores
and die.  Models of PNe formation have long been
 admittedly speculative, empirical, and  without meaningful
theoretical guidance  \citep{ib84}.  Multiple dredge-up models
reflect complex unknown stellar mixing processes.  
A counter-intuitive  1975  ``Reimer's Wind'' 
expression gives stellar mass loss rates $\dot{M} \sim  LR/M$ inversely 
proportional to the mass {M} of the star dumping its mass, where  {L} is its huge luminosity
(up to $10^{6} L_{\sun}$) and 
 {R} is its huge radius (up to $10^{2} R_{\sun}$).  Why inversely?    ``Superwinds'' 
 must be postulated  ($\S 2.3$)
to carry away unexpectedly massive stellar envelopes of surprising composition
 by forces unknown to fluid mechanics and
physics in unexpectedly dense fragments in 
standard numerical models of star evolution  \citep{pax04}.  Massive $\ge 2 M_{\sun}$
main sequence stars with mass inferred from 
gravitational-cloud-collapse luminosity-models  \citep{ibe65}
rather than JPP brightness are highly questionable.   Mass and species balances
in star, PNe and supernova formation models without HGD are uniformly problematic \citep{shi90}.

Radiation pressure, even with dust and pulsation enhancement,
 is inadequate to explain the AGB superwind  \citep{w6}.  Shock wave
effects cannot explain the large densities of the Helix cometary 
knots ($\S 3$).  All these surprises vanish when 
one recognizes that the interstellar medium consists of primordial H-He planets rather than
 a hard vacuum.     From HGD,  planetary nebulae contain $\ge 1000 M_{\sun}$
 of unevaporated JPPs from observed PNe radii $3 \times 10^{16}$ m assuming the BDM density in 
 star forming regions is
 $3 \times 10^{-17}$ kg m$^{-3}$.  Less than 1\% of these JPPs must be evaporated and
  ionized to form the unexpectedly massive stellar 
  envelopes in place rather than ejected as superwinds.  Why is it credible
  that a star can dump 94$\%$ of its mass into the ISM when it forms a carbon core?
  From HGD, the AGB-envelope-superwind concepts are failed working hypotheses like 
   CDM, $\Lambda$ and dark energy.   

 By coincidence, the direction opposite to the peak Leonid meteoroid flux
in November 2002 matched that of the closest planetary nebula (PNe) Helix (NGC 7293),
so that the Hubble Helix team of volunteers could devote a substantial
fraction of the 14 hour Leonid stand-down period taking photographs with
the full array of HST cameras, including the newly installed wide angle
Advanced Camera for Surveys (ACS).  A composite image was constructed
with a 4 m telescope ground based image mosaic  \citep{ode04} to show the
complete system.   Helix is only 219 (198-246) pc $ \approx 6.6 \times 10^{18}$ m from earth \citep{har07}
with one of the hottest and most massive known central 
white dwarf stars (120,000 K, $M_{WD} \approx M_{\sun}$), and is also the dimmest PNe \citep{gor97}.
With either a close (dMe) X-ray companion
\citep{gue01} or just a JPP accretion disk \citep{su07} it powerfully
 beams radiation and plasma into the
interstellar medium (ISM) surroundings.    Thus Helix provides an
ideal laboratory to test our claims from theory and observation 
\citep{gib96, sch96} that both the ISM of star forming
regions of galaxies and the baryonic-dark-matter (BDM) of the
universe are dominated by dense collections of volatile primordial frozen-gas planets.  

In  $\S 3$ we compare
HST/ACS Helix and other PNe observations with
HGD   and standard explanations of
cometary globule and planetary nebula formation.  Planetary nebulae from
HGD are not just transient  gas clouds emitted by dying
stars, but baryonic dark matter brought out of cold storage.   
A new interpretation of Oort cloud comets and the Oort cloud itself
appears naturally, along with evidence \citep{mat99} of Oort comet
deflection by an
$\approx 3
\times 10^{27}$ kg solar system $\ge$Jupiter-mass JPP at the Oort cavity distance $\approx 3
\times 10^{15}$ m.  The first direct detection of PFPs in Helix \citep{ode07} 
is discussed in $\S 4$ and the detection of 40,000 infrared Helix JPP 
atmospheres \citep{hor06} in $\S 5$, consistent with pulsar evidence ($\S 4,\S 5$).  

In the following $\S 2$ we  review hydro-gravitational-dynamics theory and
some of the supporting evidence, and compare the PNe predictions of HGD
with standard PNe models and the observations in $\S 3$.  We discuss
our ``nonlinear grey dust'' alternative to ``dark energy'' in $\S
4$ and summarize results in $\S 5$.  Finally, in
$\S 6$, some conclusions are offered.

\section{Theory}

\subsection {HGD structure formation}

Standard CDMHC cosmologies are based on flawed concepts about
 turbulence, ill-posed, over-simplified fluid mechanical equations, an inappropriate 
assumption that primordial astrophysical fluids are
collisionless, and the assumption of zero density to achieve a solution of the equations.  
This obsolete Jeans 1902 theory neglects non-acoustic density fluctuations, viscous forces,
turbulence forces,  particle collisions, differences between particles,
 and the effects of multiparticle mixture diffusion on
gravitational structure formation, all of which can be crucially important
in some circumstances where astrophysical structures form by self gravity.
Jeans did linear perturbation stability analysis (neglecting turbulence) of
Euler's equations (neglecting viscous forces) for a completely  uniform
ideal gas with density
$\rho$ only a function of pressure (the barotropic assumption) to reduce
the problem of self-gravitational instability to one of gravitational
acoustics.    Diffusivity effects were not considered.

To satisfy Poisson's equation $\nabla ^2 \phi = 4 \pi G \rho$ for the gravitational potential
$\phi$ of a collisionless ideal gas, Jeans assumed the density $\rho$ was zero in a
maneuver appropriately known as the ``Jeans swindle''.    The only critical
wave length for gravitational instability with all these questionable
assumptions is the Jeans acoustical length scale
$L_J$ where
\begin{equation}  
L_J \equiv V_S/(\rho G)^{1/2}
\gg (p/\rho^2 G)^{1/2} \equiv L_{JHS},
\label{eq1}
\end{equation}
   $G$ is Newton's gravitational constant and
$V_S \approx (p/\rho)^{1/2}$ is the sound speed.

The Jeans hydrostatic length scale $L_{JHS} \equiv
(p/\rho^2G)^{1/2}$  in Eq. 1 has been
misinterpreted by Jeans 1902 and others as an indication that
pressure can somehow prevent the formation of structures by gravity at length scales
smaller than $L_J$.  Viscosity, turbulence and diffusivity
can prevent small scale gravitational structure formation at Schwarz scales
(Table 1).  Pressure cannot.  In a hydrodynamic description, 
ratio $h = p/\rho$  is the stagnation specific enthalpy for  gravitational
condensation and rarefaction streamlines.  The appropriate reference 
enthalpy $h_0$ is  zero from Bernoulli's equation
$B=p/\rho + v^2 /2 = constant$ from the first law of thermodynamics for
adiabatic, isentropic, ideal gas flows at the beginning of structure
 formation, where $v=0$ is the fluid speed.  In an expanding
 universe where $v=r \gamma$ the positive rate of strain $\gamma$ is
 important at large radial values $r$ and favors fragmentation.  In the initial
 stages of gravitational instability, pressure is a slave to  the velocity
 and is irrelevant because it drops out of the momentum equation. 
 ``Where the speed is greatest the pressure is least''
 with $B$ constant quotes the usual statement of Bernoulli's law.  For
 supersonic real gases and plasmas at later stages the specific  
 enthalpy term $p/\rho$ acquires a factor of $\approx$ 5/2 famously
 neglected by Newton in his studies of acoustics without the second 
 law of thermodynamics  \citep{pil07}.  Turbulence concepts
of HGD are necessary  \citep{gib07}.

Pressure support and thermal support are concepts relevant only
to hydrostatics.  For hydrodynamics, where the velocity
is non-zero, pressure appears
in the Navier-Stokes momentum equation only in the $\nabla B \approx 0$ term.
Non-acoustic density extrema are absolutely unstable to gravitational
structure formation \citep{gib96,gib00}.  
Minima trigger  voids and maxima trigger condensates
at all scales not stabilized by turbulent forces, viscous forces, other
forces, or diffusion (see Eqs. 3-5 and Table 1 below).   The Jeans acoustic scale
$L_J$ is the size for which pressure can equilibrate acoustically without
temperature change in an ideal gas undergoing self gravitational collapse
or void formation, smoothing away all pressure forces and all pressure
resistance to self gravity.  The Jeans hydrostatic scale
$L_{JHS}$ is the size of a fluid blob for which irreversibilities such as
frictional forces or thermonuclear heating have achieved a hydrostatic
equilibrium between pressure and gravitation in a proto-Jovian-planet or
proto-star.
$L_{JHS}$ is generically much smaller than
$L_J$ and has no physical significance until gravitational condensation has
actually occurred and a hydrostatic equilibrium has been achieved.

When gas
condenses on a non-acoustic density maximum due to self gravity a variety
of results are possible.  If the amount is much larger than 
the Eddington limit 110 $M_\odot$ permitted by radiation pressure,
 a turbulent maelstrom,
superstar, and possibly  a black hole 
(or magnetosphere eternally collapsing object MECO) 
may appear.  If the amount is small, a gas
planet can form in hydrostatic equilibrium as the
gravitational potential energy is converted to heat by turbulent friction, and is
 radiated.  The pressure force
$F_P
\approx p \times L^2$ matches the gravitational force of the 
planet at $F_G \approx
\rho^2 G L^4$ at the hydrostatic Jeans scale
$L_{JHS}$.  Pressure $p$ is determined by a complex mass-momentum-energy
balance of the fluid flow and ambient conditions.  A gas with uniform
density is absolutely unstable to self gravitational structure formation on
non-acoustic density perturbations at scales larger and smaller than $L_J$
and is unstable to acoustical density fluctuations on scales larger than
$L_J$
\cite{gib96}. Pressure and temperature
cannot prevent structure formation on scales larger or smaller than
$L_J$.  Numerical simulations showing sub-Jeans scale instabilities are
rejected as ``artificial fragmentation'' based on Jeans'
misconceptions \citep{tru97}.  The fragmentation is real, and the rejection is a serious mistake.

Density fluctuations in fluids are not barotropic  as assumed by Jeans 1902
except rarely in small regions for short times near powerful sound
sources.  Density fluctuations that triggered the first gravitational
structures in the primordial fluids of interest were likely non-acoustic
(non-barotropic) density variations from turbulent mixing of temperature and
chemical species  concentrations reflecting big bang turbulence patterns
(Gibson 2001, 2004, 2005) as shown by turbulence signatures \cite{bs02,ber06} in the cosmic
microwave background temperature anisotropies.  From Jeans' theory without
Jeans' swindle, a gravitational condensation on an acoustical density
maximum rapidly becomes a non-acoustical density maximum  because the
gravitationally accreted mass retains the (zero) momentum of the motionless
ambient gas.  The Jeans 1902 analysis was ill posed because it failed to
include non-acoustic density variations as an initial condition.

Fluids with non-acoustic density fluctuations are continuously in a state
of structure formation due to self gravity unless prevented by diffusion or
fluid forces
\cite{gib96}.  Turbulence or viscous forces can dominate gravitational
forces at small distances from a point of maximum or minimum density to
prevent gravitational structure formation, but gravitational forces will
dominate turbulent or viscous forces at larger distances to cause
structures if the gas or plasma does not diffuse away faster than it can
condense or rarify due to gravity.  The concepts of pressure
support and thermal support are artifacts of the erroneous Jeans criterion
for gravitational instability.  Pressure forces could not prevent
gravitational structure formation in the plasma epoch because pressures
equilibrate in time periods smaller that the gravitational free fall time
$(\rho G)^{-1/2}$ on length scales smaller than the Jeans scale $L_J$, and
$L_J$ in the primordial plasma was larger than the Hubble scale of
causal connection $L_J > L_H = ct$, where $c$ is light speed and $t$ is
time.  Therefore, if gravitational forces exceed viscous and turbulence
forces in the plasma epoch at Schwarz scales $L_{ST}$ and $L_{SV}$ smaller
than
$L_H$ (Table 1) then gravitational structures will develop, independent of
the Jeans criterion.  Only a very large diffusivity ($D_B$) could interfere
with structure formation in the plasma.  Diffusion prevents gravitational
clumping of the non-baryonic dark matter (cold or hot) in the plasma epoch
because
$D_{NB} \gg D_B$ and $(L_{SD})_{NB} \gg L_H$.  Diffusion
does not prevent fragmentation of the baryonic material at 30,000 years when
$(L_{SD})_{B} \le L_H$.

Consider the gravitational response of a large motionless body of
uniform density gas to a sudden change at time  $t=0$ on scale
$L \ll L_J$ of a rigid mass perturbation $M(t)$ at the center, either a
cannonball or vacuum beach ball depending on whether $M(0)$ is positive
or negative \citep{gib00}. Gravitational forces cause all the surrounding gas to
accelerate slowly toward or away from the central mass perturbation. 
 Integrating the radial gravitational acceleration $dv_r/dt = -GM/r^2$ gives the radial velocity
\begin {equation}
v_r = -GM(t)t r^{-2}
\end {equation}
so the central mass increases or decreases at a rate 
\begin {equation}
dM(t)/dt = -v_r 4 \pi r^2 \rho =
4 \pi \rho G M(t)t
\end {equation} substituting the expression for $v_r$.  
Separating variables and integrating gives 
\begin {equation} 
M(t) =M(0) exp(\pm 2 \pi \rho G t^2), t \ll t_{G} 
\end {equation}
respectively, where nothing much happens for time periods less
than the gravitational free fall time $t_G = (\rho G)^{-1/2}$ except for a
gradual build up or depletion of the gas near the center.  Note that the classic
Bondi-Hoyle-Littleton accretion rate $dM_{BH}/dt \approx 4 \pi \rho [GM]^2 V_S^{-3} $ \citep{kru06, pad05, edg04}
on point masses $M$ in a gas of density $\rho$ is contradicted.  The sound
speed $V_S$ is irrelevant to point mass accretion rates for the same reasons $V_S$
is irrelevant to gravitational structure formation for times $t \ll t_G$. 
  
For condensation, at
$t = 0.43 t_G$ the mass ratio $M(t)/M(0)$ for $r < L$ has increased by only
a factor of 2.7, but goes from 534 at $t = t_G$ to  $10^{11} $ at
$t = 2 t_G$ during the time it would take for an acoustic signal to reach a
distance $L_J$.  Hydrostatic pressure changes are concentrated at the Jeans hydrostatic
scale $L_{JHS} \ll L_J$.  Pressure support and the Jeans 1902 criterion clearly
fail in this exercise, indicating failure of CDMHC cosmology.

The diffusion velocity is
$D/L$ for diffusivity
$D$ at distance $L$ \citep{gib68a,gib68b} and the gravitational velocity is $L (\rho G)^{1/2}$.
The two velocities are equal at the diffusive Schwarz length scale
\begin{equation} L_{SD} \equiv [D^2 / \rho G]^{1/4}.\end{equation}  Weakly
collisional particles such as the hypothetical cold-dark-matter (CDM)
material cannot possibly form clumps, seeds, halos, or  potential wells for
baryonic matter collection because the CDM particles have large diffusivity
and will disperse, consistent with observations \cite{sa02}.  Diffusivity
$D
\approx V_p \times L_c$, where $V_p$ is the particle speed and $L_c$ is the
collision distance. Because weakly collisional particles have large
collision distances with large diffusive Schwarz lengths the non-baryonic
dark matter (possibly neutrinos) is the last material to fragment by self
gravity and not the first as assumed by CDM cosmologies.  The first
structures occur as proto-supercluster-voids in the baryonic plasma
controlled by viscous and weak turbulence forces, independent of
diffusivity  ($D \approx \nu$).  The CDM seeds postulated as the basis of
CDMHCC never happened because $(L_{SD})_{NB}
\gg ct$ in the plasma epoch.  Because CDM seeds and halos never happened,
hierarchical clustering of CDM halos to form galaxies and their clusters
never happened (Gibson 1996, 2000, 2001, 2004, 2005, 2006a, 2006b).

Cold dark matter was invented to explain the observation that gravitational
structure formed early in the universe that should not be there from the
Jeans 1902 criterion that forbids structure in the baryonic plasma because
$(L_J)_B > L_H$ during the plasma epoch (where sound speed approached light
speed
$V_S =c/\surd 3$). In this erroneous CDM cosmology, non-baryonic particles with rest
mass sufficient to be non-relativistic at their time of decoupling are
considered ``cold'' dark matter, and are assumed to form permanent, cohesive
clumps in virial equilibrium that can only interact with matter and other
CDM clumps gravitationally.  This assumption that CDM clumps are cohesive
is unnecessary, unrealistic, and fluid mechanically untenable.  Such clumps
are unstable to tidal forces because they lack particle collisions necessary
 to produce  cohesive forces to hold them together  \citep{gib06a}.  

Numerical simulations
of large numbers of falsely cohesive CDM clumps  show a tendency for the
clumps to clump further due to gravity to form ``dark matter halos'', falsely justifying the
cold dark matter hierarchical clustering cosmology (CDMHCC).  The
clustering ``halos'' grow to $10^6 M_\sun$ by about $z=20$ \citep{abe02} as the
universe expands and cools and the pre-galactic clumps cluster.   Gradually the
baryonic matter (at $10^{16}$ s) falls into the growing gravitational potential wells of
the CDM halos, cools off sufficiently to form the first (very massive and very
 late at 300 Myr) Population III stars 
whose powerful supernovas reionized all the gas of the universe \citep{she06}.  
However, observations
show this never happened \citep{ah06,gib06a}.  Pop-III photons are not detected, consistent with
the HGD prediction that they never existed and that the first stars formed at $10^{13}$ s (0.3 Myr) 
and were quite small except at the cores of PGCs at the cores of the protogalaxies.  
The missing hydrogen cited as (``Gunn-Peterson trough'') 
evidence for reionization is actually sequestered as PGC 
clumps of frozen JPP planets.  As
we have seen, CDMHCC is not necessary since the Jeans 1902 criterion is
incorrect.  Baryons (plasma) begin gravitational structure formation during the
plasma epoch when the horizon scale exceeds the largest Schwarz scale \citep{gib96,gib00}.  

Clumps of collisionless or collisional CDM would either form black holes 
or thermalize
in time periods of order the gravitational free fall time $(\rho G)^{-1/2}$
because the particles would gravitate to the center of the clump by core
collapse where the density would exponentiate, causing double and triple
gravitational interactions or particle collisions that would thermalize the
velocity distribution and trigger diffusional evaporation.  For collisional
CDM, consider a spherical clump of perfectly cold CDM with mass
$M$, density
$\rho$, particle mass $m$ and collision cross section $\sigma$.  The clump
collapses in time $(\rho G)^{-1/2}$ to density $\rho_c =
(m/\sigma)^{3/2}M^{-1/2}$ where collisions begin and the velocity
distribution thermalizes.  Particles with velocities greater than the
escape velocity $v \approx 2MG/r$ then diffuse away from the clump,
where
$r=(M/\rho)^{1/3}$ is the initial clump size.  For typically
considered CDM clumps of mass $\approx 10^{36}$ kg and CDM particles more
massive than
$10^{-24}$ kg (WIMPs with $\sigma \approx 10^{-42}$ m$^2$ small
enough to escape detection) the density from the expression would require a
collision scale smaller than the clump Schwarzschild radius so that such
CDM clumps would collapse to form black holes.  Less massive motionless CDM
particles collapse to diffusive densities smaller than the
black hole density, have collisions, thermalize, and diffuse away.  From the
outer halo radius size measured for galaxy cluster halos it is possible to
estimate the non-baryonic dark matter particle mass to be of order
$10^{-35}$ kg (10 ev) and the diffusivity to be $\approx 10^{30}$ m$^2$
s$^{-1}$
\citep{gib00}.  Thus, CDM clumps are neither necessary nor physically
possible, and are ruled out by observations
\citep{sa02}.  It is  recommended that the CDMHC scenario for
structure formation and cosmology be abandoned.

The baryonic matter is subject to large viscous forces, especially in the
hot primordial plasma and gas states  existing when most gravitational
structures first formed \citep{gib00}.  The viscous forces per unit  volume
$\rho
\nu
\gamma L^2$ dominate gravitational forces $\rho^2 G L^4$ at small scales,
where
$\nu$ is the kinematic viscosity and $\gamma$ is the rate of strain of the
fluid.  The forces match at the viscous Schwarz length
\begin{equation} L_{SV} \equiv (\nu \gamma /
\rho G)^{1/2},\end{equation}
   which is the smallest size for self gravitational condensation or void
formation in such a flow.  Turbulent forces may permit larger mass
gravitational structures to develop; for example, in thermonuclear
maelstroms at galaxy cores to form central black holes.  Turbulent forces
$\rho
\varepsilon^{2/3} L^{8/3}$ match gravitational forces  $\rho^2 G L^4$ at the turbulent
Schwarz scale
\begin{equation}L_{ST} \equiv \varepsilon ^{1/2}/(\rho
G)^{3/4},\end{equation}
   where $\varepsilon$ is the viscous dissipation rate of the turbulence.
Because in the primordial plasma the viscosity and diffusivity are identical
and the rate-of-strain $\gamma$ is larger than the free-fall frequency
$(\rho G)^{1/2}$, the viscous and turbulent Schwarz scales
$L_{SV}$ and
$L_{ST}$ will be larger than the diffusive Schwarz scale $L_{SD}$, from
(3), (4) and (5).

The criterion for structure formation in the plasma epoch is that
both $L_{SV}$ and $L_{ST}$ become less than the horizon scale $L_H = ct$.
Reynolds numbers in the plasma epoch were near critical, with  $L_{SV}
\approx L_{ST}$.  From $L_{SV}< ct$, gravitational structures first formed
when
$\nu$ first decreased to values less than radiation dominated values $c^2 t
$ at time 30,000 years or
$t
\approx 10^{12}$ seconds \cite{gib96}, well before $10^{13}$ seconds which
is the time of plasma to gas transition (300,000 years).  Because the
expansion of the universe inhibited condensation but enhanced void
formation in the weakly turbulent plasma, the first structures were
proto-supercluster-voids in the baryonic plasma. At
$10^{12}$ s
\begin{equation} (L_{SD})_{NB} \gg L_{SV} \approx L_{ST} \approx 5 \times
L_K \approx L_H = 3
\times 10^{20} \rm m \gg (L_{SD})_B, \end{equation}  where $(L_{SD})_{NB}$
refers to the non-baryonic component and $L_{SV}$,
$L_{ST}$, $L_{K}$, and $(L_{SD})_B$ scales refer to the baryonic (plasma)
component.  Acoustic peaks inferred from CMB
spectra  reflect acoustic signatures of the first gravitational
structure formation and the sizes of the voids (see $\S4$ Fig. 9).  These
supercluster voids are cold spots on the CMB and completely
empty at $10^{25}$ m scales from radio telescope measurements \citep{rud07}.
Because such scales require impossible clustering
speeds, this strongly contradicts $\Lambda$CDMHCC.

As proto-supercluster mass plasma fragments formed, the voids filled with
non-baryonic matter by diffusion, thus inhibiting further structure
formation by decreasing the gravitational driving force.  The baryonic mass
density
$\rho
\approx 3
\times  10^{-17}$ kg/$\rm m^3$ and rate of strain
$  \gamma \approx 10^{-12}$ $\rm s^{-1}$ were preserved as hydrodynamic
fossils within the proto-supercluster fragments and within proto-cluster
and proto-galaxy objects resulting from subsequent fragmentation as the
photon viscosity and
$L_{SV}$ decreased prior to the plasma-gas transition and photon decoupling
\cite{gib00}.  As shown in Eq. 6, the Kolmogorov scale $L_K \equiv [\nu^3
/\varepsilon ]^{1/4}$ and the viscous and turbulent Schwarz scales at the
time of first structure matched the horizon scale $L_H
\equiv ct \approx 3 \times 10^{20}$ m, freezing in the density,
strain-rate, and spin magnitudes and directions of the subsequent
proto-cluster and proto-galaxy fragments of proto-superclusters.  Remnants
of the strain-rate and spin magnitudes and directions of the weak
turbulence at the time of first structure formation are forms of fossil
vorticity turbulence
\cite{gib99}.

The quiet condition of the primordial gas is  revealed by measurements of
temperature fluctuations of the cosmic microwave background  radiation that
show an average $\delta T/T \approx 10^{-5}$ too small for much
turbulence to have existed at that time of plasma-gas transition ($10^{13}$
s).  Turbulent plasma motions were strongly damped by buoyancy forces at
horizon scales after the first gravitational fragmentation time
$10^{12}$ s.  Viscous forces in the plasma are inadequate to explain the
lack of primordial turbulence ($\nu$
$ \ge 10^{30}$ m$^2$ s$^{-1}$ is required but, after $10^{12}$ s, $\nu \le 4
\times 10^{26}$, Gibson 2000). The observed lack of plasma turbulence
proves that large scale buoyancy forces, and therefore self gravitational
structure formation, must have begun in the plasma epoch $\approx
10^{11} \-- 10^{13}$ s.

The gas temperature, density, viscosity, and rate of strain are all
precisely known at transition, so the gas viscous Schwarz  mass
$L_{SV}^3 \rho$ is
$10^{24-25}$ kg, the mass of a small planet (Mars-Earth), or about
$10^{-6} M_{\sun}$, with uncertainty a factor of ten.  From HGD, soon
after the cooling primordial plasma turned to gas at $10^{13}$ s (300,000
yr), the entire baryonic universe condensed to a fog of hot planetary-mass
primordial-fog-particle (PFPs) clouds, preventing collapse at the accoustic Jeans mass.
In the cooling universe these gas-cloud objects cooled and shrank, formed H-He rain, and
froze solid to become the BDM and the basic material of
construction for stars and everything else, presently $\approx 30 \times 10^{6}$ rogue
planets per star in trillion-planet
Jeans-mass ($10^{36}$ kg) PGC clumps.

The Jeans mass $L_J^3 \rho$ of the primordial gas at transition was about
$10^6 M_{\sun}$, also with $\approx \times 10$ uncertainty, the mass of a
globular-star-cluster (GC).  Proto-galaxies fragmented at the PFP scale but
also at this proto-globular-star-cluster PGC scale
$L_J$, although not for the reasons given by the Jeans 1902 theory.  Density
fluctuations in the gaseous proto-galaxies were absolutely unstable to
void formation at all scales larger than the viscous Schwarz scale
$L_{SV}$.  Pressure can  only remain in equilibrium with density without
temperature changes in a gravitationally expanding void on scales smaller
than the Jeans scale.  From the second law of thermodynamics, rarefaction
wave speeds are limited to speeds less than the sonic velocity. 
Density  minima expand due to gravity to form
voids subsonically.  Cooling could
therefore occur and be compensated by radiation in the otherwise isothermal
primordial gas when the expanding voids approached the Jeans scale.
Gravitational fragmentations of  proto-galaxies were then accelerated by
radiative heat transfer to these cooler regions, resulting in fragmentation
at the Jeans scale and isolation of proto-globular-star-clusters (PGCs)
with the primordial-gas-Jeans-mass. 

These PGC objects were not able to collapse from their own self
gravity  because of their internal fragmentation at the viscous Schwarz
scale to form $\approx 10^{24}$ kg PFPs. The fact that globular star clusters have
precisely the same density $\approx \rho_{0}$ and
primordial-gas-Jeans-mass from galaxy to galaxy proves they were all
formed simultaneously soon after the time of the plasma to gas
transition $10^{13}$ s.  The gas has never been so uniform since, and no
mechanism exists to recover such a high density, let alone such a high
uniform density, as the fossil turbulent density value
$\rho_{0} \approx 3 \times 10^{-17}$ kg/$\rm m^3$.  Young globular
cluster formation in BDM halos in the Tadpole, Mice, and Antennae galaxy
mergers
\citep{gs03a} show that dark PGC clusters of PFPs are remarkably stable
structures, persisting without disruption or star formation for more than
ten billion years.

\subsection{Observational evidence for PGCs and PFPs}

Searches for point mass objects as the dark matter by looking
for microlensing of stars in the bulge and the Magellanic clouds
produced a few reliable detections and many self-lenses, variable stars and background
supernova events, leading to claims by the MACHO/OGLE/EROS consortia that this
form of dark matter has been observationally excluded \citep{alc98}.
These studies have all assumed a uniform (``Gaussian'') density rather than the 
highly clumped
(``log-Gaussian'') density \citep{gs99} with a non-linear frictional accretion cascade
 for the MAssive Compact Halo Objects (MACHOs) expected from HGD.  Sparse sampling 
 reduces detection sensitivity to small clumped planetary
mass objects.  Since  PFPs within PGC clumps must
accretionally cascade over a million-fold mass range to produce JPPs and
stars their statistical distribution becomes an intermittent lognormal that will
profoundly affect an appropriate  sampling strategy and microlensing data
interpretation.  This rules out the exclusion of PFP mass objects as the baryonic
dark matter (BDM) of the Galaxy by MACHO/OGLE/EROS
\citep{gs99}.  OGLE campaigns focusing on large
planetary mass ($10^{-3} M_{\sun}$) to brown dwarf  mass 
objects have revealed 121 transiting
and orbiting candidates, some with orbits less than one day \citep{uda03}.  
  More recent
observations toward M31 give 95\% confidence level claims
that brown dwarf MACHOs comprise $\approx$20\%
of the combined MW-M31 dark matter halo  \citep{cal05}. 
Estimates for the MW
halo of $\approx 0.2 M_\sun$ lenses from LMC stars have 
increased to $\approx$16\% \citep{alc00,ben05}. 
Both of these estimates increase by a large factors ($\gg 5$)
from HGD when JPP clumping into PGCs  is taken into account since
microlensing by a JPP planet requires a line of sight passing through a PGC and
PGCs (as CHVCs) occupy a small fraction of the sky ($\approx 1\%$) 
with HVC wakes ($\approx 20\%$).

Evidence that planetary mass objects dominate the BDM in galaxies has been
gradually accumulating and has been reviewed \citep{gs03b}.
Cometary knot candidates for PFPs and JPPs appear whenever hot events
like white dwarfs, novas, plasma jets, Herbig-Haro objects, and supernovas
happen, consistent with the prediction of HGD that the knots reveal Jovian
planets that comprise the BDM, as we see for the planetary nebulae in the
present paper.  However, the most convincing evidence for our hypothesis,
because it averages the dark matter over much larger volumes of space, is
provided by one of the most technically challenging areas in astronomy; that
is, quasar microlensing
\citep{sch96}.  Several years and many dedicated observers were required to
confirm the Schild 1996 measured time delay of the Q0957 lensed quasar images so
that the twinkling of the subtracted light curves could be confirmed and the
frequency of twinkling interpreted as evidence that the dominant point mass
objects of the lensing galaxy were of small planetary mass.  

By using
multiple observatories around the Earth it has now been possible to
accurately establish the Q0957 time delay at
$417.09  \pm 0.07$ days (Colley et al. 2002, 2003).  With this
unprecedented accuracy a statistically significant microlensing event of
only 12 hours has now been detected
\citep{ColS03} indicating a $7.4
\times 10^{22}$ kg (moon-mass) PFP.  An additional microlensing system has been
observed (Schechter et al. 2003) and confirmed, and its time delay measured
(Ofek and Maoz 2003).  To attribute the microlensing to stars rather than
planets required Schechter et al. 2003 to propose relativistic knots in
the quasar.  An additional four lensed quasar systems with  measured time
delays show monthly period microlensing.   These studies 
support the prediction of HGD that  the masses of their galaxy lenses are dominated 
 by small planetary mass objects as the baryonic dark matter
 (Burud et al. 2000, 2002; Hjorth et al. 2002) that may produce intermittent
 systematic dimming errors rather than dark energy \citep{sd06}.

Flux anomalies in four-quasar-image gravitational lenses have 
been interpreted as
evidence \citep{dal02}  for the dark matter substructure predicted by CDM halo 
models, but the anomalies may also be taken as evidence for
concentrations of baryonic dark matter such as PGCs, especially when the
images are found to twinkle with frequencies consistent with the existence
of planetary mass objects.  Evidence that the small planetary objects
causing high frequency quasar image twinkling are clumped as PGCs is indicated
by the HE1104 \citep{sche03} damped Lyman alpha
lensing system (DLA $\equiv$ neutral hydrogen column density larger than
$10^{24.3}$ m$^{-2}$), suggesting PGC candidates from the
evidence of gas and planets.  Active searches are underway for quasar lensing DLAs
with planetary frequency twinkling that can add to this evidence of PGCs.  
Twenty $10^5 - 10^6 M_{\sun}$ (PGC-PFP) galaxy halo objects have been detected
$\ge 10^{21}$ m  from M31 \citep{thi04}. 

Perhaps the most irrefutable evidence for galaxy inner halos
 of baryonic PGC-PFP clumps is the HST/ACS image
showing an aligned row of $42 \-- 46$ YGCs (see $\S 2.3.2$ and Fig. 1) precisely tracking
the frictionally merging galaxy fragments VVcdef in the Tadpole system 
\citep{gs03a}.  Concepts of collisionless fluid mechanics and collisionless
tidal tails applied to merging
galaxy systems are rendered obsolete by this image.
Numerous YGCs are also seen in the fragmenting galaxy cluster Stephan's
Quintet-HGC 92 \citep{gs03c}.  The mysterious red shifts of this dense 
Hickson Compact Galaxy Cluster (HGC)
support the HGD model of sticky beginnings of the cluster in the plasma
epoch, where viscous forces of the baryonic dark matter halo of the cluster
have inhibited the final breakup due to the expansion of the universe to
about 200 million years ago and reduced the transverse velocities of the
galaxies to small values so that they appear aligned in a thin pencil by
perspective.  Close alignments of QSOs with bright galaxies (suggesting
intrinsic red shifts) have been noted for many years \citep{hoy00}, but are
 easily explained by the HGD concept that proto-galaxies formed
 in the plasma epoch by viscous-gravitational fragmentation
 of larger  objects termed proto-galaxy-clusters \citep{gib00}.

\subsection{Planetary Nebula formation}
\subsubsection{The standard model}

According to the standard model of white dwarf and planetary nebula formation, an ordinary
star like the sun burns less than half of its hydrogen and helium to form a
hot, dense, carbon core \citep{bus99,ib84}.   White dwarf masses are typically
$0.6 M_{\sun}$ or less even though the initial star mass is estimated
from the increased brightness at WD formation to be $8 M_{\sun}$ or more.  Are
intermediate stars really so massive?  If so, how does all this mass escape?  Radiation pressures
are much too small for such massive ejections either as winds, plasma beams, or clumps, even
assisted by dust and pulsations \citep{w6}.    Most of the large mass inferred from the brightness probably just reflects the bright JPP atmospheres as the massive frozen planets  near the new 
white-hot dwarf evaporate and ionize, incleasing the JPP entrainment rate by friction.
Claims of red-blue supergiant stripping \citep{mau04} are highly questionable.  
The brightness of the JPP atmospheres 
masquerades as huge central stellar masses and envelopes.
Friction from the gas and dust of evaporated JPPs accelerates the formation of the PNe and the growth of central carbon white dwarf(s) toward either a SNe Ia event or a bypass of the event by enhanced
 mixing of the carbon core giving critical mass $1.4 M_{\sun}$ iron-nickel cores and SNe II events.
 How much of the $9-25 M_{\sun}$ mass of SNe II remnants can be attributed to the precursor star?

 Interpretation of nuclear chemistry from spectral results to describe the physical processes
 of stellar evolution to form white dwarfs  \citep{her05} is limited by a poor understanding of modern
 stratified turbulent mixing physics.  Methods that
 account for fossil and zombie turbulence radial internal wave
 transport in mixing chimneys are required \citep{kee05, gib06c, gib06d, gib07}
focusing on the smallest scales \citep{wan06}.  New information about carbon stars
 is available at the critical infrared spectral bands of 
 cool AGB stars from the $Spitzer$ $Space$ $Telescope$ \citep{lag06} but the mass
 loss problem remains unsolved.  Crucial contributions of 
 mass and luminosity from the ISM are not
 taken into account in the standard models of PNe formation and evolution and in standard models
 of star formation and evolution.

From standard star models, the neutral atmosphere of a dying red giant with approximate density
$\rho \approx 10^{-17}$ kg m$^{-3}$ \citep{cha01} is somehow expelled 
to the ISM along with a very massive
(but unobserved and likely mythical) envelope by 
(unexplained) dynamical
and photon pressures when the hot,
$T \approx 10^{5}$ K, dense, $\rho \approx 10^{10}$ kg m$^{-3}$, carbon
core is exposed as a white dwarf star with no source of fuel unless
accompanied by a donor companion.  The density of this
$10^{16}$ kg atmosphere expanded to the distance of  the inner Helix
radius is trivial ($\approx 10^{-29}$ kg m$^{-3}$).   
At most this could bring the PNe ejected atmosphere
density to a small fraction ($\approx 10^{-15}$) of
$\rho \approx 10^{-14}$ kg m$^{-3}$ values observed in the knots
\citep{mea98}.  Why are small and intermediate mass
main sequence stars  ($1-9 M_{\sun}$) so inefficient
that they burn only a small fraction of their initial mass before
they die to form  ($0.5-1.44 M_{\sun}$) white dwarfs?  We suggest
small stars are likely to be more efficient than large stars in their burning
of gravitationally collected mass.  What they don't burn is returned as helium
and carbon white dwarfs, neutron stars (if the small star
gets large) and SNe ashes, not superwinds.  The ashes
are collected gravitationally by the $\ge 1000 M_{\sun}$ of ambient PFPs and JPPs 
influenced by an average star in its evolution from birth to death, and a small fraction returned to
the stars as the dust of comets.  From HGD, most intermediate mass stars and superwinds are 
obsolete working hypotheses used to explain unexpected 
brightness of JPP atmospheres formed from the ISM and not ejected  \citep{her05}.

From radio telescope 
measurements \citep{kna82} large stars up to $9
M_{\sun}$ form white dwarfs and companions with huge envelopes that
 have complex histories with superwind
Asymptotic Giant Branch (AGB) periods where most of the assumed initial mass
of the star  is mysteriously
expelled into the ISM
\citep{bus99}.  The possibility is not mentioned in the literature
 that the ISM itself could be supplying the unexpectedly large, luminous,
envelope masses and superwind mass losses
 inferred from radio and infrared telescope measurements \citep{kna82}, OH/IR 
stars \citep{deJ83}, and 
star cluster models  \citep{cla01}.  
It has been speculated in versions of the standard model
that shock wave instabilities somehow produce cometary knots 
ejected by PNe central stars  \citep{vis94, vis83},
 or that a fast wind impacts the
photo-ionized inner surface of the dense ejected envelope giving
Rayleigh-Taylor instabilities that somehow produce the cometary globules and radial wakes
observed
\citep{gar06, cap73}.  Such models produce cometary globule densities  much smaller than
observed, and require globule wake densities much larger than observed.

Several other problems exist for standard  PNe models
 without HGD.  Huge ($3-9 M_{\sun}$) H-He masses observed in PNe  are richer in
other species and dust than one would expect to be expelled as stellar
winds or cometary bullets during any efficient solar mass star evolution, where most
of the star's H-He fuel should presumably be converted by thermonuclear fusion to
carbon in the core before the star dies.  More than a solar mass
of gas and dust is found in the inner nebular ring of Helix, with a dusty
H-He-O-N-CO composition matching that of the interstellar medium rather
than winds from the  hydrogen-depleted atmosphere of a carbon
star, but up to $11 M_\sun$ may be inferred for the total PNe
\citep{spe02}.  The cometary globules are too massive and too dense to
match any Rayleigh-Taylor instability model.  Such models \citep{gar06} give
cometary globule densities of only $\rho \approx 10^{-19}$ kg m$^{-3}$ compared
to $\rho \approx 10^{-14}$ kg m$^{-3}$ observed.   

The closest AGB C star is IRC+10216 \citep{mau99}.  It is brighter than any star at long
wavelengths, but invisible in the blue from strong dust absorption.  Loss rates inferred
from its brightness are large ($2 \times 10^{-5}{M_{\sun}}$/yr).  
The possibility that the mass indicated by the brightness
could have been brought out of the dark in place has not been considered.  Multiple, fragmented
and asymmetric rings are observed, indicating a central binary.  The rings
are irregular and extend to $4 \times 10^{15}$ m, with central brightness of the envelope
confined to $2 \times 10^{14}$ m.  The observed ring structures appear to be
 wakes of Jovian orbital planets evaporating in response to the 
red giant growth and powerful radiation from the central star(s).  Rather
than superwinds outward we see the effects of enhanced JPP accretion inward, clearing
the Oort cloud cavity prior to PNe formation.

The density increase
due to maximum expected Mach 6 hypersonic shock waves in astrophysical gases is only
about a factor of six, not $10^{5}$.  Rayleigh-Taylor instability,
where a low density fluid accelerates a high density fluid, causes
little change in the densities of the two fluids. Turbulence
dispersion of nonlinear thin shell instabilities
\citep{vis94, vis83} should decrease or prevent shock induced or
gravitational increases in density.  The masses of the inner Helix
cometary globules 
 are measured and modeled to be $\gg 10^{25}$
kg, much larger than expected for PFP planets that have not merged with other PFPs to form
globulette clumps and JPPs.  
No mechanism is known by which such massive dense objects can form
or exist near the central star.  Neither could they be ejected
without disruption to the distances where they are observed.
Measurements of proper motions of the cometary knots provide a
definitive test of whether the knots are in the gas and expanding at the
outflow velocity away from the central binary, as expected
in the standard model, or moving 
randomly with some collapse component toward the center.  Proper
motion measurements to date
\citep{ode02} suggest they are mostly moving 
randomly with approximately virial PGC speeds (also see Fig. 2 below in $\S 3$).

\subsubsection{The HGD model}

According to HGD, all stars are formed by accretion of PFP planets,
larger Jovian PFP planets (JPPs), and brown and red dwarf stars
within a primordial PGC interstellar medium.  The accretion mechanism is likely to be
binary with clumping, where two JPPs experience a near collision so that internal tidal forces and
frictional heating of their atmospheres produces evaporation of the frozen H-He
planets and an increase in the amount of gas in their atmospheres.  Smaller JPP and PFP
planets within the atmospheres are collected as comets or merging moons.  
Increased size and density of planet atmospheres
from collisions, tidal forces, or star radiation results in ``frictional hardening'' of binary planets
until they fragment, evaporate, merge and then shrink and refreeze by radiation.
  The binary accretion cascade to larger mass clumps of planet-binaries and star-binaries
continues until inhibited by thermonuclear processes.  
Hence ``3 out of every 2 stars is
a binary'' (personal communication to RES from Cecilia Helena Payne-Gaposchkin, pioneer astronomer).  This classic astronomical overstatement
could actually be true if one of the ``stars'' is a binary and the other a binary
of binaries or a triplet
of two binaries and a rogue.  Small binary stars with lots of moons and planets 
is the signature of star formation
from planets.   Large single stars with no planets and no moons is what you get from the large
clouds of gas collected by CDM halos \citep{abe02,she06}.

Exotic clumped binary-star and binary-planet systems are highly likely from the 
non-linear nature of HGD star and JPP planet formation.  Heating from
a binary PFP merger results in a large atmosphere for the double-mass PFP-binary 
that will increase its cross section for capture of more PFPs in growing clumps.  
Evidence is accumulating
that most PNe central stars are binaries as expected from HGD \citep{dem04,sok06,moe06}.  One
of the brightest stars in the sky is Gamma Velorum in Vela, with two binaries and two rogues all
within $10^{16}$ m of each other, the nominal size of a PNe.  One of the binaries is a WR star and
a blue supergiant B-star with $1.5 \times 10^{11}$ m (1 AU) separation.  From
aperture masking interferometry using the 10-m aperture Keck I telescope, 
another WR-B binary is the
pinwheel nebula Wolf-Rayet 104 in Sagitarrius, where the stars are separated by only 
 $3 \times 10^{11}$ m (3 AU) and are  $10^5$ brighter than the sun  \citep{tut99}.  How much
 of the apparent brightness and apparent masses of these systems is provided by
 evaporating JPPs?  The pinwheel nebula dust clouds are surprising this close to large
 hot stars ($\approx 50$kK) that should reduce dust to atoms.  Complex 
 shock cooling induced dust models from colliding superwinds  \citep{uso91,pil07}
 to explain the dust of pinweel nebulae are unnecessary if the stars are
 accreting a rain of dusty evaporating JPPs.  When one member of the binary dies to
 become a shrinking white-dwarf it appears that the smaller star begins to eat the larger 
 one  (creating a brown dwarf desert).  
 Few WD binaries are found with large mass ratios  \citep{hoa07}. 
 Either the WD binary has an equal mass companion or just a JPP accretion disk \citep{su07}.

Large PFP atmospheres from mergers and close encounters  increase their
frictional interaction with other randomly encountered ambient PFP atmospheres. 
This slows the relative motion of the objects and increases the time between their
collisions and mergers.  Radiation to outer space will cause the PFP
atmospheres to cool and eventually rain out and freeze if no further
close encounters or collisions occur, leading to a new state of metastable equilibrium with the
ambient gas.  To reach Jupiter mass,
$10^{-6} M_{\sun}$ mass PFPs and their growing sons and daughters must pair
10 times ($2^{10} \approx 10^{3}$). To reach stellar mass, 20 PFP binary
pairings are required ($2^{20} \approx 10^{6}$).  Because of the binary
nature of PFP structure formation through JPPs, it is clear that matched double stars 
and complex systems will result, as observed, and that the stars will have 
large numbers of large gassy planets with many moons that the stars 
capture in orbit or absorb as comets, as observed.  

Rocky and nickel-iron cores of planets like the Earth and  rock-crusted
 stainless-steel Mercury in this scenario are simply the rocky and 
iron-nickel cores of rogue interstellar
Jupiters that have processed the SiO dust, water, organics, iron and nickel particles
 accumulated gravitationally from
supernova remnants in their cores and in the cores of the thousands of
PFPs that they have accreted to achieve their masses.  Rather than being accreted
as comets by the growing star, these massive JPPs were captured in orbits and their
gas layers evaporated as their orbits decayed to leave terrestrial planets \citep{vit06}.  They
were not created by any star as often assumed \citep{bos95,bos04}.  A hot-Saturn exoplanet
observed in orbit at only $6 \times 10^{9}$ m with a $4 \times 10^{26}$ kg rocky-metal core 
conflicts with such stellar Jeans gravitational instability models  \citep{sat05} as well as with
core accretion-gas capture models \citep{kor02} and proto-star dynamical fragmentation
models \citep{bod00}.  From HGD this size core implies a multi-Jupiter
or red dwarf that has already lost most of its atmosphere to the central star or stars.

Without PFPs, the
existence of rocks and unoxidized iron-nickel cores of planets are mysteries.  It has been
known since the end of the bronze age that very high temperatures are
needed for carbon to reduce iron oxides to metalic iron, as will happen in
supernovas IIab where hydrogen, silicon and carbon will form oxides by 
reducing oxides of  iron and nickel to metal particles.  All this stardust will be swept
up by gravitational fields of the PFPs, which should by now
 be deeply crusted with dry magnetic talcum powder
after their $\approx$13 billion years in existence as interstellar gravitational 
vacuum cleaners.  Samples
of cometary material confirm large quantities of stardust in 
comet tails and in comet bodies.  Crashing a 364 kg object into
Jupiter comet Tempel 1 revealed a $\gg 10$m deep crust of 1-100 micron particle size low strength
powder  \citep{ahe05,fel06,sun06}.  Exotic refractory dust particles lacking
carbonates or hydrates as collected by 
NASA's  $Stardust$ mission to Jupiter comet  Wild 2  require high temperatures of
formation ($\gg$800-1400 K) and no wetness  \citep{bro06}.
Unexpectedly powerful mixing from creation in sub-Mercury orbits with
transport far beyond Jupiter must be postulated 
for such comets to be created as the sun formed.  
The comets are easy to understand as pieces of PFP crust  
if stars, large planets and comets all form from dusty primordial planets
in primordial PGC molecular clouds.

Large gas planets from PFP accretion cascades may form gently
over long periods, with ample time at every stage for
their atmospheres to readjust with ambient conditions and return to
metastable states of random motion.  These are probably the conditions under
which the old globular star clusters (OGCs) in the halo of the Milky Way
Galaxy formed their small long-lived stars, and their large ancient planets
\citep{sig03}.  However, if the PFP accretional cascade is forced by
radiation or tidal forces from passing stars or ambient supernovae, a
more rapid cascade will occur where the PFP atmospheres become large and
the relative motions in a PGC become highly turbulent.  
The turbulence will mix
the PFPs and their large planet descendants and inhibit large average
density increases or decreases.  In this case another instability becomes
possible; that is, if the turbulence weakens the creation of large
central density structures but enhances accretion and clumping to form large planets
and brown dwarfs, the increase of density at an accretion center can become so rapid that
buoyancy forces may develop from the density gradients.  This will
suddenly damp the turbulence at the Schwarz turbulence scale $L_{ST}$
(see Table 1) to produce fossil turbulence
\citep{gib99} in a volume containing $> M_{\sun}$ of gas, PFPs,
JPPs, and a complex of binary stars in formation.

Turbulence fossilization due to buoyancy then creates a
gravitational collapse of an accretion center of the resulting non-turbulent gas and
PFPs from the sudden lack of turbulence resistance.  A fossil turbulence
hole in the ISM will be left with size determined by the turbulence levels
existing at the beginning of fossilization.  The accretion of the planets
and gas within the hole will be accelerated by the rapidly increasing
density.  The total mass of the stars produced will be the volume of the 
``Oort cloud'' hole (Oort cavity) times the ISM density.  If the mass is many solar masses then the
superstars formed will soon explode as supernovae, triggering a sequence of
ambient PFP evaporations, accretional cascades, and a  starburst that may
consume the entire dark PGC and its PFPs to produce a million stars and a young
globular cluster (YGC) or a super-star cluster \citep{tra03}.  

Numerous YCCs
are triggered into star formation by galaxy mergers, such as the merging of two
galaxies and some fragments revealing a 130 kpc ($4
\times 10^{21}$ m) radius baryonic dark matter halo in the  VV29abcdef
Tadpole complex imaged by HST/ACS immediately after installation of the ACS
camera \citep{ben04}.  Figure 1 shows an SSC dwarf galaxy, revealed in the baryonic dark
matter halo of the central Tadpole galaxy VV29a by a dense narrow trail
of YGCs pointing precisely to the spiral star wake produced as the dwarf
blue galaxy VV29c and companions VV29def merged with VV29a \citep{gs03a}.

Planetary nebulae form when a small star formed by gradual PFP accretion
uses up all its H-He fuel to form a dense white dwarf carbon core with temperature less
than $8 \times 10^{8}$ K.  The high exposed  core
temperature and spin enhanced radiation of plasma in jets
and winds from the white dwarf 
evaporate JPPs remaining nearby from its red giant phase
giving an AGB (asymptotic giant branch) red giant star (eg: Arcturus)
that appears to be a massive envelope from its brightness.
  As discussed
previously for the standard model, red
giant stars have envelope diameter
$\approx 10^{12}$ m, atmosphere density $\approx 10^{-17}$ kg m$^{-3}$
and a
$6 \times 10^6$ m diameter
$\rho \approx 10^{10}$ kg m$^{-3}$ carbon star core
\citep{cha01}.   The total mass expelled is thus only $\approx 10^{16}$ kg, or
$\approx 10^{-8} M_{\sun}$, much less than the gas mass values $(3-1)
\times  M_{\sun}$ claimed to be observed in planetary nebulae
from their luminosity.  Without HGD, this much
gas is mysterious.

Because white dwarfs have close companion stars
and continuous JPP accretion, as observed 
for Helix and Cats-Eye  and easily inferred for many
other PNe's \citep{ode02},  the companions and JPP accretion disk will
 contribute to the WD growth.  Spinning magnetic field lines at the white dwarf
poles and magnetohydrodynamic turbulence at its equator capture the incoming plasma 
and magnetic fields to produce powerful plasma jets, plasma winds, and photon radiation. 
The accretion disk of the white dwarf may shield some of its radiation and broadly
beam some of its radiation.  The following observations show the effects of such plasma 
beams and radiation in PNe.

\section{Observations}

Figure 2 is an HST image of the Large Magellanic Cloud PNe N66 (SMP 83, WS 35, LM1-52) 
at a distance of $1.7 \times 10^{21}$ m (57 pc), taken on 06/26/1991 with the European
Space Agency Faint Object Camera (FOC)  filtered for
540 seconds at the 5007 
$\rm \AA$ doubly ionized oxygen emission line (O III).  This remarkable object is the only
confirmed PNe where the central star is classified as a Wolf-Rayet of the nitrogen 
sequence type (WN).  LMC-N66 has recently exhibited highly variable brightness, with an indicated
mass loss rate increase from 1983 to 1995 by a factor of 40 \citep{pen97}.
 Its central binary is surrounded by a looped $\approx 6 M_{\sun}$ mass
  PNe of  bright cometary globules, 
 interpreted in Fig. 2 as partially evaporated clumping JPPs 
 in multi-Jupiter mass globulettes \citep{gah07}.  From spectal analysis and modeling the
  central binary system is
a $1.2 M_{\sun}$ white dwarf  with a non-degenerate companion that is
building the WD toward the Chandrasekhar limit and a SN Ia event in 
$\approx 10^{5}$ years \citep{pen04, ham03} along with
the rain of JPP comets.  Part of the mass stream to the WD
is ejected as a nutating plasma beam that evaporates JPPs in the
 looped arcs shown in Fig. 2.  The mass $M_{PNe}$ of the observed
 nebular material is more than $5 M_{\sun}$.
 From HGD this implies about 0.5$\%$ of the $1000 M_{\sun}$ PFPs and JPPs of the
 interstellar medium within the $2.5 \times 10^{16}$ m  luminous 
 range of the PNe have been brought out
 of the dark  by radiation and plasma jets from the central star system. 
 
 Some of the JPPs in Fig. 2 have
detectable O III emission (5007 $\rm \AA$) wakes that indicate JPP velocities $V_{JPP}$ are in 
random directions with at least virial values, as expected from HGD.  The virial velocity
$V_{vir} = (2MG/r)^{1/2}$ for a  PGC is about $1.7 \times 10^{4}$ m/s, where $M$ 
is the PGC mass, $G$ is
  Newton's constant, and $r$ is the PGC radius.  In PGC metastable equilibrium the PFP speed should 
  slow to less than $V_{vir}$ due to gas friction, so that the growth of JPPs and the rate of star 
  formation are low.  Most PGCs will never develop a star.  Some have developed a million.
 
 Wolf-Rayet (WR) stars are very bright, red, and until recently have been claimed to be very massive 
($\ge 20 M_{\sun}$) with large mass loss rates ($\ge 10^{-5} M_{\sun}$/yr).  Star mass models
derived from the increased luminosity with mass of gas cloud collapse \citep{ibe65} are unreliable
if stars form from planets.  
WR stars are often found with
 surrounding nebulae \citep{mor03}, generally in galaxy disks where
 PGCs are accreted and where $V_{JPP}$ values should be large.  High He, C, N, and O concentrations
 suggest final stages of evolution toward white dwarf status for at least one of the 
 central stars.  HST images reveal most WRs to be binary or in multiple star systems,
 with numerous dense clumps in their envelopes that appear to be evaporating globulette-JPPs
 as in Fig. 2.   Most of their claimed mass and most of their claimed
  large mass loss rates are likely the result of 
 bright evaporating JPPs spin-radiated by a central dying star and misinterpreted as massive 
 stellar envelopes and superwinds.  For example, see WR124 in nebula M1-67, STSci-1998-38 in the
 HST archives.    From the 1998 news release nebular clumps have 
 mass $2 \times 10^{26}$ kg and scale $10^{14}$ m, giving
 a density  $10^{-16}$ kg m$^{-3}$.   How can such massive dense objects be ejected from a star?
 From HGD the clumps are clearly bright JPP planet atmospheres evaporated from the PGC-ISM.  
 Vast overestimates result for WR star masses, SN II precursor star masses \citep{mau04, p0d93}, and initial White-Dwarf star masses.  To explain the
 extreme brightness of SN 1993J with a single star without JPPs requires a  $40 M_{\sun}$ precursor
   mass \citep{ald94}.

Figure 3 shows a standard  stellar-model white dwarf mass evolution 
diagram for planetary
nebulae  in Praesepe (circles) and Hyades (squares) star clusters  \citep{cla01}, compared to Helix and LMC-N66 and our pulsar precursor model.
In an 80 PNe collection \citep{gor97}, Helix has the most massive central white dwarf.  It is also
the dimmest, and therefore has the smallest estimated
PNe mass  $M_{PNe}$  (shown as an open star in Fig. 3).  From HGD,  
  $M_{Initial} \approx M_{PNe}$ masses are vastly 
  overestimated from the brightness of evaporated JPPs that dominate  $M_{PNe}$.     
  Observed $M_{PNe}$
should therefore not be interpreted as the initial 
masses  $M_{Initial}$ of central white dwarfs, as assumed using
the standard PNe model \citep{wei00}.   Masses of WR stars and SNe II 
precursors are overestimated
from their extreme JPP atmosphere brightness and implied 
super-massive Hayashi tracks \citep{ibe65} by even 
larger factors.  Rather than 
$9-25 M_\sun$ \citep{gel07, mau04, shi90} such stars are likely no larger than $1.4 M_\sun$ since neutron stars have mass $1.4 M_\sun$ \citep{tho99} precisely known from pulsar timing. 
Any central WD of a PNe has the possibility to grow to the SNe Ia size by
accretion of JPPs, as shown by the LMC-N66 (hexagon) point in Fig. 3.  If the WD is
fed by a companion red giant (RB)  accretion disk \citep{hac01} as well as by JPP rain the
probability of a SN Ia or SN II event increases.  In the final
stages of growth, the WD will likely be surrounded by a PNe similar to that of Helix, 
where the central Oort cavity sphere has been depleted of JPPs by gravity to form the central star and
its JPP accretion disk \citep{su07} and the PNe is formed
by radiative evaporation of JPPs to form large atmospheres by plasma jets
and plasma and photon radiation powered by JPP accretion and
the rapid spinning and complex magnetohydrodynamics of the 
central white dwarf.  Supernova Ia events will always  be
subject to intermittent dimming depending on the line of sight.  PNe can appear around hot
central stars at any time during the life of the star, which is more than $10^{10}$ years for the
small stars leading to SNe Ia events, not $10^{4}$ years as assumed in the standard PNe model.

Figure 4 shows a mosaic of nine HST/ACS images from the F658N filter
(H$_{\alpha}$ and N II) that enhances the  ionized cometary
globules and their hydrogen tails (http://archive.stsci.edu 
 /hst/helix/images.html).    A sphere with radius
$5 \times 10^{15}$ m is shown.  A much smaller sphere will give an adequate supply of
primordial-fog-particles (PFPs) with PGC primordial 
mass density $ \rho_0 = 3 \times 10^{-17}$ kg
m$^{-3}$  to form two central solar mass stars.  
The large comets closest to the central stars must be evaporating
massive planets (Jupiters) to survive measured evaporation rates of
$2 \times 10^{-8} M_{\sun}$  year$^{-1}$
\citep{mea98} for the 20,000 year kinematic age of Helix.  Massive planets are
formed in the accretional cascade of PFPs to form stars according to HGD.
The younger (2,000 year old) planetary nebula Spirograph (IC 418) shown
below shows
shock wave patterns from the supersonic stellar winds but no cometary PFP
candidates within its fossil turbulence accretion sphere corresponding
to the Oort cloud source of long period comets.  

Is the sun surrounded by 
an Oort cavity of size $L_{Oort}\approx (M_{star}/\rho_{ISM})^{1/3} \approx  4\times10^{15}$ m
 reflecting its $1 M_{\sun}$ of accretion
 in a PFP dominated interstellar medium with primordial
 density $\rho_0$?  Rather
than an Oort cloud of comets, are long period ``Oort Cloud Comets''
actually PFP and JPP planets accreting from the
 inner boundary of the Oort cavity?  In a remarkable
application of celestial mechanics to Oort cloud comets (``Cometary
evidence of a massive body in the outer Oort cloud'')
 a  multi-Jupiter (JPP) mass perturber within 5$\degr$ of the Galactic plane
has been inferred \citep{mat99}.  Out of 82 ``new'' first-time entrant long
period comets with orbit scales less than $10^{16}$ m, 29 have 
aphelion directions on the same great circle, suggesting that the
galactic-tide-Saturn-Jupiter loss cylinder has been smeared inward along
the track of the perturber.  An apparently independent $99.9\%$
detection 
\citep{mur99} gives the object a retrograde orbit with period of $5.8
\times 10^6$ years assuming it is gravitationally bound to the sun, and
excludes a variety of explanations (eg: star encounter, solar system
ejection) for its existence as extremely unlikely.  The object is easy
to explain from HGD  as one of many JPPs at distance 
$\ge 2 \times 10^{15}$ m on the
inner surface or our Oort cavity as in Figs. 4  for Helix, Fig. 7 for
Eskimo and Fig. 8 for Dumbbell PNes. 

Assuming density $\rho_0$, the  inner  spherical nebular shell for Helix
contains
$\approx 20 M_{\sun}$ of dark PFPs, from which $ 1.5
\times M_{\sun}$ has been evaporated as gas and dust
\citep{spe02}.
Evidence for bipolar beamed radiation is shown by the brighter regions of
the nebula in Fig. 4 at angles 10 and 4 o'clock, and by the light to dark
transition after 11:30 suggesting the bipolar beam is rotating slowly
clockwise.  Note that the tails of the comets are long ($\approx 10^{15}$
m) before 11:30 and short or nonexistent afterward.  Rayleigh-Taylor
instability as a mechanism to produce the globules
\citep{cap73} gives densities much too
low.  The WD plasma beams appear to have started rotation at about 1 o'clock
with deep penetration of the radiation on both sides, revolved
once, and is observed with  bright edge at 11:30 having completed less
than two revolutions to form the Helix spiral.

Figure 5 shows a Hubble Space Telescope  Helix WFPC2 1996 image to the
northeast in Helix where the closest comets to the center are found
\citep{ode96}.  The cometary globules have size about
$6 \times 10^{13}$ m and measured atmospheric 
mass $(2-11) \times 10^{25}$ kg \citep{mea92, ode96,  hug02}, with
spacing $\approx 3 \times 10^{14}$ m, as expected for 
$\approx 8 \times 10^{26}$ kg  evaporating JPP gas planets 
 in a relic concentration corresponding to the primordial
($\rho_0$) plasma density
$3 \times 10^{-17}$ kg m$^{-3}$ at the time of first
structure 30,000 years after the big bang.  These largest cometary
globules must have much larger mass planets than PFPs at their cores to have
survived the 20,000 year lifetime of the Helix planetary nebula
with measured mass loss rates of order
$10^{-8} M_{\sun}$ year$^{-1}$ \citep{mea98}.  
  The spacing of the cometary knots becomes closer for
distances farther from the central stars, consistent with these objects
having PFPs or small JPPs at their cores (see Fig. 11 below).

Figure 6 shows an example of the new ACS/WFPC composite images from the
northern region of Helix nebula confirming the uniform density of the cometary
globules.  From HGD this reflects the uniform 
ambient distribution of virialized, dark-matter,
frozen PFPs and JPPs  expected in PGCs with primordial
density $ \rho_0 \approx 3 \times 10^{-17}$ kg m$^{-3}$.   Thus planets 
provide raw material
to produce and grow the central white dwarf of a PNe by binary 
hierarchical clustering and clumping.
Once clustering and clumping begins the star formation is highly non-linear.  The more clustering
the more gas.  The more gas the more clustering.  
Thus the  larger JPPs and globulettes of PFPs in the Jupiter mass range
are more likely to be found at Oort cavity distances
 and the brown dwarfs near the center or merged as stars and close binaries.  
Within the ionized cavity of the PNe accreted JPPs  are
evaporated by radiation from the white dwarf and absorbed or captured in orbits.  
Images and properties of ``knots'' in Helix, Eskimo,
Dumbbell and other PNe
\citep{ode02} are consistent with this PFP-JPP-globulette star formation interpretation.

Figure 7 shows the Eskimo planetary nebula (NGC 2392), which 
at $7.2 \times 10^{19}$ m (2.4 kpc) is 11 times
more distant from earth than Helix \citep{gor97},
but is still close enough for numerous cometary
globules to be resolved by HST cameras.  The PNe is smaller than Helix and has a central
shocked region with no comets, just like the small, even younger, Spirograph nebula shown at the
bottom of Fig. 4.   Eskimo PNe has a few very large widely separated
cometary globules dominating its brightness, suggesting these may be evaporating 
JPPs with multi-Jupiter masses.  Note the large gas wakes without cometary globules
at the 6 o'clock position (in the beard).  Presumably these were even brighter while
their JPPs were evaporating.

Figure 8 shows details of the central region of the Dumbbell planetary
nebula featuring numerous cometary globules and knots.  The spacing of the
objects is consistent with PFP and JPP planets with average mass density
$\approx 10^4$ times the average $10^{-21}$ kg m$^{-3}$ for the Galaxy.  Because
of their primordial origin, planets with this same density $\rho_0$ dominate the mass and
species content of the ISM in all galaxies, fossilizing the primordial baryonic density  from
the time of first structure in the plasma epoch 30,000 years after the big
bang as predicted by HGD.  The dumbbell morphology reflects the existence of a binary
central star system and its plasma beam jet to evaporate JPPs at the Oort cavity edge.

\section{The ``Nonlinear Grey Dust'' Systematic Error of ``Dark Energy''}

Figure 9 summarizes the hydro-gravitational-dynamics (HGD) theory
of gravitational structure formation leading
to the formation of baryonic dark matter and the resulting
 dark energy misconception \citep{gib05}.  A Planck
scale quantum-gravitational instability triggers big bang turbulent 
combustion.  The resulting
turbulent temperature patterns are fossilized by nucleosynthesis in 
the energy epoch  as random H-He density fluctuations, which seed the first
gravitational formation of structure by fragmentation at the horizon and Schwarz viscous
scale and Schwarz turbulent scale in the plasma epoch
 ($L_H \approx L_{SV} \approx L_{ST}$, Table 1).   The first
gravitational structures are super-cluster-voids starting at $10^{12}$
seconds and growing with sonic ($c/3^{1/2}$) speed until the plasma-gas transition at
$10^{13}$ s (300,000 years).  An observed -73 $\mu$K CMB cold spot
reflects a $10^{25}$ m void for $z \le 1$ \citep{rud07} as expected from HGD. 
 The maximum probability
for such a large void from $\Lambda$CDMHCC 
is  much less than $10^{-9}$  \citep{hoy04}.

The smallest gravitational fragments from the
plasma epoch are proto-galaxies formed by fragmentation
with the $10^{20}$ m Nomura scale  
($L_N \approx L_{ST} \ge L_{SV} \approx L_K$, Table 1) and chain-clump
spiral-clump morphology \citep{gib06b}
along stretching turbulent vortex lines and compressing spirals of the weak plasma
 turbulence.  The kinematic viscosity reduction by a factor of $10^{13}$ gives 
 two fragmentation scales and structures in the primordial 
 gas; that is, the $10^6 M_{\sun} $ Jeans acoustic scale and 
PGCs, and the $10^{-6} M_{\sun} $ viscous Schwarz scale and the PFPs.   With time the
planetary mass PFPs freeze to form the baryonic dark matter. 
Some small fraction accrete to form JPPs and stars.  Because SNe Ia
occur surrounded by evaporating PFPs and JPPs, a random
``nonlinear grey dust'' dimming of the supernova brightness is likely.

Figure 10 shows the proposed interpretation of the ground based (open
circle) and
\citep{rie04} HST/ACS (solid circle) SNe Ia dimming for red shifts $ 0.01
\le z
\le 2$.    The wide scatter of the amount of SNe Ia dimming appears to be real,
and is consistent with our ``nonlinear grey dust'' model, where a random
amount of absorption should be expected depending on the line of sight to
the supernova and the degree of evaporation of the baryonic dark matter
interstellar medium.  A ``uniform grey dust'' systematic dimming increases
with $z$ contrary to observations.  The ``dark
energy'' concept seems unlikely because it requires a radical change in
the physical theory of gravity.  HGD requires changes in the standard
(CDMHCC) model of gravitational structure formation and the interpretation
of planetary nebulae.  The slight random dimming found in the observations
is just what one expects from ``nonlinear grey dust'' formed as the growing
hot carbon star evaporates Oort cavity and rim planets to form
large, cold, dusty atmospheres that may be on the line of sight to the SNe
Ia that eventually occurs.  Planetary nebulae such as the Helix and other
PNe described above illustrate the process we are suggesting.  Radiation from the
proto-SNe Ia can be directly from the shrinking carbon star or from plasma jets and winds
formed as any companion stars or just a JPP accretion disk \citep{su07}  feeds its carbon growth.  

Figure 11 shows a closeup view of clumped JPPs and PFPs in the northern rim of the Helix Oort cavity
imaged in molecular hydrogen $\rm H_2$ at 2.12 microns \citep{ode07}.  The distance scales
are derived from the new O'Dell et al.  images and a trigonometric parallax 
estimate of 219 (198-246) pc for the distance to Helix \citep{har07}.  O'Dell et al. 2007 conclude
 $\rm H_2$ is in local thermodynamic equilibrium, as expected for evaporated planet atmospheres
produced by intense radiation and not by any shock phenomenon.

Figure 12 shows a collection of interstellar medium electron density spectral estimates often referred to as the ``Great Power Law on the Sky''  from the remarkable agreement with the same $q^{-11/3}$ Kolmogorov, Corrsin, Obukhov spectral form over 11 decades of wavenumber $q$
 \citep{arm81, gib91, arm95}.  An `inner scale' at $\approx 10^{12}$ m from
 pulsar scintillations can be understood
 as the Obukhov-Corrsin scale
  $L_C \equiv (D/\varepsilon)^{1/4}$ marking the beginning of a fossil turbulence inertial-diffusive
 spectral subrange $q^{-15/3}$ (shown in Fig. 12), where electron density is a strongly diffusive passive
 scalar property with diffusity $D$ $\approx 30 \times$ larger than the kinematic viscosity $\nu$ \citep{gib68a,gib68b,gib07}.  Evidence that observed pulsar scintillation spectra \citep{you07} represent forward scattering from
  discrete features is provided by observations
 of parabolic scintillation arcs \citep{tra07, hil05}.  The observed 
 scintillations with small inner scales 
 $\le 10^{10}$ m and possibly  $\le 10^{7}$ m indicate turbulent partially ionized
   atmospheres evaporated from dark matter
 planets by neutron star supernova and  pulsar plasma jets and radiation. The
 $\varepsilon \approx 1$ m$^2$ s$^{-3}$ and Reynolds number $\approx 10^5$ values
 implied by the spectra
 indicate fossilized strong turbulent mixing driven by the planet evaporation.
 For a gas to be a fluid and produce the
  observed scintillations with strong turbulence and turbulent mixing spectra requires
 gas density $\rho \ge 10^{-14}$ kg m$^{-3}$  in
 the scintillation regions; that is,
 higher than average in the Galaxy by $\ge 10^7$.  Such a large density matches that
 measured for the cometary globules of Helix, also interpreted as dark matter planet atmospheres.
 The well-studied Crab supernova II remnant at $6 \times 10^{19}$ m shows strong evidence 
 of evaporated dark matter planets with wakes
 pointing to the path of the pulsar and its close companions, revealing powerful pulsar jets
 and evaporated JPP atmospheres in a series of HST images.  Atmosphere
 diameters (1 arc sec `knots') as large as $3 \times 10^{14}$ m are 
 reported \citep{sch02}, $\approx \times 10$ 
 larger than those in Helix.
 
Extreme scattering events (ESE), multiple images of radio pulsars and quasars, 
and other indications of small $\approx 10^{13}$ m dense 
$\approx 10^{-12}$ kg m$^{-3}$ refractors
are detected frequently using the $\le 10^{-3}$ 
arc sec resolution of radio telescopes.  From the detection frequency
and the size and $\approx 10^{27}$ kg mass of the ionized and neutral clouds
 observed,  authors have suggested 
 the objects may provide most of 
the  Galaxy mass \citep{wal98, hil05}.  We agree.  From HGD,  these
radio telescope detections manifest gas atmospheres of
 Galaxy disk  primordial dark matter Jovian planets.   No other explanation
 seems plausible.

\section{Discussion of results}

HGD theory combined with high-resolution multi-frequency space telescope observations
 require major changes in the standard models of cosmology, star
formation,  star death, and planetary nebulae formation.  
Trails of young globular clusters in merging galaxies and bright gases 
of clumpy  nebulae and dense masers surrounding
white-dwarf stars, Wolf-Rayet stars and neutron stars provide a growing body of clear evidence
for the existence of the millions of frozen primordial planets (PFPs and JPPs) per star in 
metastable-planet-clusters (PGCs) suggested \citep{gib96, sch96} as the 
baryonic dark matter and interstellar medium.   Even though the primordial planets
are dark and distant they make their presence known because of their enormous total mass, 
the brightness and scattering properties of their evaporated atmospheres, and
 because they are the raw material for everything else.  PFPs are now individually detectable 
 in the Helix PNe from 
 their $\rm H_2$ signal at 2.12 microns (Fig. 11).  JPPs with molecular atmospheres
  in great abundance ($\ge$ 20,000-40,000) are detected
 in Helix from their 5.8 and 8 micron purely rotational
  lines of  molecular hydrogen  from the infrared array camera
 IRAC on the $Spitzer$ space telescope with Helix 
 resolution of $6 \times 10^{13}$ m  \citep{hor06}.   Since these JPP atmospheres are mostly 
 evaporated by spin-powered beamed plasma jets and winds of the white dwarf,
  they represent a small fraction of the $> 1000 M_{\sun}$ 
 of JPPs in its range.  Similar results are obtained from the NICMOS NIC3 camera with
 the 2.12 micron  molecular hydrogen filter from HST \citep{mei05}. 

When disturbed from equilibrium by tidal forces or radiation, JPP planets grow to stellar mass
by binary accretion with neighbors where friction of close encounters causes growth
of planetary atmospheres, more friction, merger, reprocessing and cooling to a new
state of metastable equilibrium with shrinking planetary atmospheres as the gases refreeze. 
This non-linear binary cascade to larger size gas planets in pairs and pairs of pairs leads to star 
formation as stellar binaries within the PGC clumps.  It explains the presence 
of the massive JPPs that persist in Helix in the shells closest to the central stars, as 
shown in Figs. 4, 5, 6, 10 and 11.  HGD theory explains why most stars are binaries 
and why most galaxies and galaxy clusters are not.  Stars are formed by hierarchical
clustering of planets, not condensation within gas clouds, and galaxies and galaxy clusters
 are formed in strings and spirals by gravitational fragmentation of weakly turbulent 
  plasma, not by hierarchical clustering of CDM halos.  Measured supervoid sizes are so large
 they must have begun growth by gravitational fragmentation in the plasma epoch as predicted by HGD.
 The CDMHCC paradigm should be abandoned along with $\Lambda$ and dark energy.
 Supernova models based on supergiant stars and their $9-25 M_{\sun}$ envelopes 
 are highly questionable along with PNe models and estimates of WD and SNe II precursor star
 masses (Fig. 3).

Hundreds of PNe are observed in the LMC and SMC galaxies, interpreted from HGD as tidally
agitated star forming clumps of PGCs in the BDM halo of the Milky 
Way equivalent to the super-star-clusters (SSCs) of YGCs
 formed in the BDM halo of the Tadpole (VV29) merger (Fig. 1).  Fig. 1 \citep{tra03, gs03a} shows
the SSC as a linear string of $\ge 42$ young globular star clusters with star formation triggered by 
passage of one of the merging galaxy fragments (VV29cdef) passing through the BDM halo of VV29a.
The YGCs have 3-10 My ages, showing they must have been formed in place and not ejected as
a frictionless tidal tail in this 500 My old merging system.  Collisionless fluid mechanical modeling
of galactic dynamics with frictionless tidal tails instead of star trails in the baryonic dark matter
 is highly misleading and should be abandoned.

One of the LMC PNe (LMC-N66) 
has recently shown strong brightness variation and appears to be in the final
stages of white-dwarf growth leading to a SNE Ia event  \citep{pen04}.  
Fig. 2 shows an archive HST image
at the 5007 $ \rm\AA$ wavelength OIII emission line.  The heavy rain of JPPs on the
central star(s) appears to be adding carbon to the WD and fueling a powerful plasma beam
that brings JPPs and their O III wakes out of the dark  to 
diameters $\approx 5 \times 10^{16}$ m, a
region containing $\ge 1000 M_{\sun}$ of JPPs using the primordial density
 $\rho _0 \approx 3 \times 10^{-17}$ kg m$^{-3}$ from HGD.  The wakes
 point  in random directions, indicating that the evaporating JPP 
 velocities $V_{JPP}$ are large and random, consistent with a
  rapid JPP accretion rates and rapid white-dwarf growth.  Will the C-N core
  of the WD be mixed away by this rapid accretion to give Fe and a SN II event?

From HGD and the observations it seems clear that most stars form as binary
star systems from Jovian primordial planets that grow by binary accretion
within dark primordial PGC clumps of such planets.  
Unless the JPPs are strongly agitated the stars formed will be small.  
The white dwarf and its companion can then both slowly grow toward the 
 Chandrasekhar limit drawing mass from accreted JPPs and possibly each other.
Because SNe Ia events result from dying small stars that have very long
lives, we can understand why it took nearly 10 billion years 
with red shift $z=0.46$ for ``dark energy'' effects to appear, and why SNe Ia
events are not seen at red shifts much larger than 1.  

  Why do all pulsars twinkle?
  It is because the progenitors of neutron stars form in PGCs
   surrounded and generously fed by partially evaporated planets with large atmospheres.
   Over 50 pulsar binaries and doubles in complex dense systems including exo-planets have
   been detected \citep{sig03,bis06, lyn04,tho99}.
   The supernovae IIab in agitated Galaxy disk PGCs where pulsars usually occur
   generate so many large atmosphere JPPs that all lines of sight to pulsars
   pass through at least one large turbulent or previously turbulent
    planet atmosphere (Fig. 12).  It should therefore be no surprise that less strongly agitated
   PGCs forming poorly mixed carbon stars that explode as SNe Ia events will
   occasionally have unobscured and occasionally obscured lines of sight (Fig. 10),
   calling the dark energy hypothesis into question.

\section{Conclusions}

High resolution wide angle HST/ACS images and 4m ground based
telescope images \citep{ode04} confirm and extend the previous
WFPC2 HST picture of the Helix planetary nebula \citep{ode96} showing
thousands of closely-spaced cometary globules.  Slow comet rains on
central white dwarfs in dense primordial metastable molecular clouds of planets are interpreted
from HGD as  generic features of PNes.  Frozen BDM planets 
(PFPs and JPPs sometimes in globulettes)  evaporated 
by spin driven plasma beams
and plasma winds from the central white dwarf form the PNe.  
A slow rain of planet-comets grow the WD to Chandrasekhar instability.  Evaporation rates from 
the largest cometary globules suggest they possess $\ge$Jupiter 
mass cores  \citep{mea98}, consistent with background radiation 
absorption masses
 $(2-11) 10^{25}$ kg  for the Helix cometary globule atmospheres
  \citep{mea92, ode96, hug02}.  From their
 spacing and HGD, the largest cometary globules have multiple Jupiter 
 masses (Fig. 5) and are termed JPPs.  From HGD, all
 SN Ia events should occur in PNes subject to random dimming by JPP atmospheres, as
   observed (Fig. 10) and misinterpreted as dark energy.

Models are questioned for planetary nebula formation where
Rayleigh-Taylor instabilities of a postulated (unexplainably) dense and massive
superwind  outer shell are triggered by
collision with a later, rapidly expanding, less-dense, inner shell to
form the globules \citep{cap73, gar06}.  
Such two wind PNe models cannot account for the morphology, regularity, and
large observed densities and masses of the globules.   Speculations that
accretional shocks or variable radiation pressures in stars can trigger gravitational instabilities
\citep{vis94, vis83} to achieve such large density differences underestimate powerful
 turbulence, radiation, and molecular dispersion forces existing in 
 stellar conditions that would certainly smooth away any such dense globules.
 From HGD, superwinds are not necessary to explain PNes and never happen.  Star
 mass overestimates indicating superwinds have neglected the brightness
 of evaporated JPP atmospheres (Fig. 3).  Luminous galaxy masses have
 probably been overestimated for the same reason and should be reexamined.
 Stars with mass $2 M_\sun  \rightarrow 100 M_\sun$ may exist, but require strong turbulent mixing
 from exceptionally large JPP accretion rates.

No convincing mechanism exists to 
produce or expel dense objects from the central stars of PNe.  Dense 
OH and SiO maser cloudlets near red giants observed
by high resolution radio telescopes 
\citep{alc86a,alc86b} are interpreted as 
evaporating JPPs from their high densities $\rho \ge 10^{-9}$ kg m$^{-3}$.  Such densities
cannot be achieved by shocks in the relatively thin red giant 
atmospheres.  Shock fronts
can be seen to exist in younger PNe than Helix (Figures 4 and 7) but are not
accompanied by any cometary globules.  Models for white dwarf and planetary nebula
formation  \citep{bus99,ib84} cannot and do not explain either the cometary globules
or the tremendous loss of mass for stars with initial mass  $M_{Initial} = (1-9) M_{\sun}$
by superwinds to  form  white dwarfs 
with final mass   $M_{Final} = (0.5-1) M_{\sun}$ (Fig. 3).  Observations of post AGB
stars with multiple masers and bipolar flows reflect JPP evaporation and accretion, not envelope
formation \citep{zij01}.

We conclude that a better model  for interpreting the observations is
provided by hydro-gravitational-dynamics theory (HGD, Fig. 9), where the brightest
cometary globules in PNe are indeed comets formed when radiation and plasma jets
from the white dwarf and its companion evaporate and reveal volatile
frozen gas planets of the ISM at Oort cavity distances.  Observed Oort cavity sizes (Figs. 4, 7, 10)
$L_{Oort} \approx 3 \times 10^{15}$ m produced  when a
 star forms from accreting planets in a PGC confirms the 
 the primordial density $\rho_0$ of HGD for the first gravitational structures
from the expression  $L_{Oort}\approx (M_{\sun}/\rho_{0})^{1/3} = 4 \times 10^{15}$ m.
  The planets are
JPP Jovian accretions of primordial-fog-particle (PFP)  frozen
H-He  proto-planets formed at the plasma to gas transition 300,000 years after
the big bang in proto-globular-star-cluster (PGC) clumps
\citep{gib96}, consistent with quasar microlensing observations showing
a lens galaxy mass dominated by rogue planets ``likely to be the
missing mass''
\citep{sch96}.  All stars are formed from primordial planets in these dense primordial clumps.

From HGD and all observations, PFPs and
JPPs in PGCs dominate the mass and gases of the inner halo 
mass of galaxies within a radius 
of about 100 kpc ($3 \times 10^{21}$ m) as most of the dark matter. 
Proto-galaxies formed during the plasma epoch
fragmented after transition to gas at primordial Jeans and Schwarz
scales (Table 1) to form PGC clouds of
PFPs that comprise the BDM and ISM of all inner
galaxy halos.  From  HST/ACS Helix images and previous
observations, the density of the Galaxy disk ISM is that of
proto-superclusters formed 30,000 years after the big bang; that is,
$\rho_{0}
\approx 3
\times  10^{-17}$ kg/$\rm m^3$, preserved as a hydrodynamic fossil and
revealed by the
$(10-4) \times 10^{13}$ m separations of the PFP candidates (cometary
globules) observed in Helix that imply this density (Fig. 11).

HST images of other nearby planetary nebula support our interpretation.
Cometary globules brought out of the dark by beamed and other spin enhanced radiation
from a shrinking central white dwarf is a
generic  rather than transient feature of planetary nebulae.
Thus, the ISM is dominated by small frozen accreting planets
with such  small separations that the indicated mass density is that of a PGC; 
that is,  $\rho \approx \rho_{PGC} \approx \rho_0$, which is 
 $\approx 10^{4}$ larger than that of the Galaxy.  Standard PNe models that
suggest planetary nebulae are brief ($10^4$ year) puffs of star dust from dying white
dwarfs by superwinds from massive envelopes
 (Fig. 3) must be discarded.  Large PNe masses  $M_{PNe}$
formed by the evaporation of JPPs (Fig. 2) must not be confused
with  $M_{Initial}$ for the white dwarf (Fig. 3).  Infrared detections of dense
molecular hydrogen clumps in Helix  from both HST
and $Spitzer$ space telescopes provide 
$\approx$ 40,000 JPP and PFP clump (globulette) 
candidates \citep{ode07, mei05,hor06} from the
$\ge 1000 M_{\sun}$ mass of JPPs and PFPs expected to exist in the observed
$2.5 \times 10^{16}$ m radius of Helix from HGD.  JPPs should form
within globulette clumps of PFPs from the stickiness of PFP atmospheres \citep{gah07}.

From HGD, most stars form as binary pairs from the binary and globulette accretion of
baryonic dark matter PFP and JPP planets in PGC clumps leaving Oort cloud size
holes in the ISM.  When one of the stars in a binary forms a white dwarf it can draw
on the fuel of its companion and accreted JPPs
 to form a  PNe of cometary globules  (Fig. 2) and a proto-SNe
Ia from the growing central stars (Fig. 3).  Radiation from the 
pair can be seen as precessing plasma jets that
evaporate rings of cometary globules as in the Helix PNe (Figs. 4, 5, 6, 10, 11) and in
other planetary nebulae (Figs. 2, 7, 8).  These JPP atmospheres give the
``nonlinear grey dust'' random-systematic-error-dimming indicated in Fig.
10.  Pulsar scintillation spectra (Fig. 12) require stratified turbulence and turbulent mixing
in dense weakly ionized gases \citep{gib07}, strongly indicating large evaporated
Jovian planet atmospheres agitated by the supernova and pulsar winds
 as seen in HST images of the Crab nebula and inferred
 for PNe in Fig. 10.  Both the dark energy concept and the 
$\Lambda$CDM cosmology required to justify dark energy seem hopelessly problematic.

\acknowledgments

Critically important information in this paper would not be available without the
heroic work and dedication of astronaut John Mace Grunsfeld whose
amazing preparation and skills in the fourth space telescope
repair mission made HST/ACS images possible.

\clearpage

\begin{figure}
         \epsscale{.7}
         \plotone{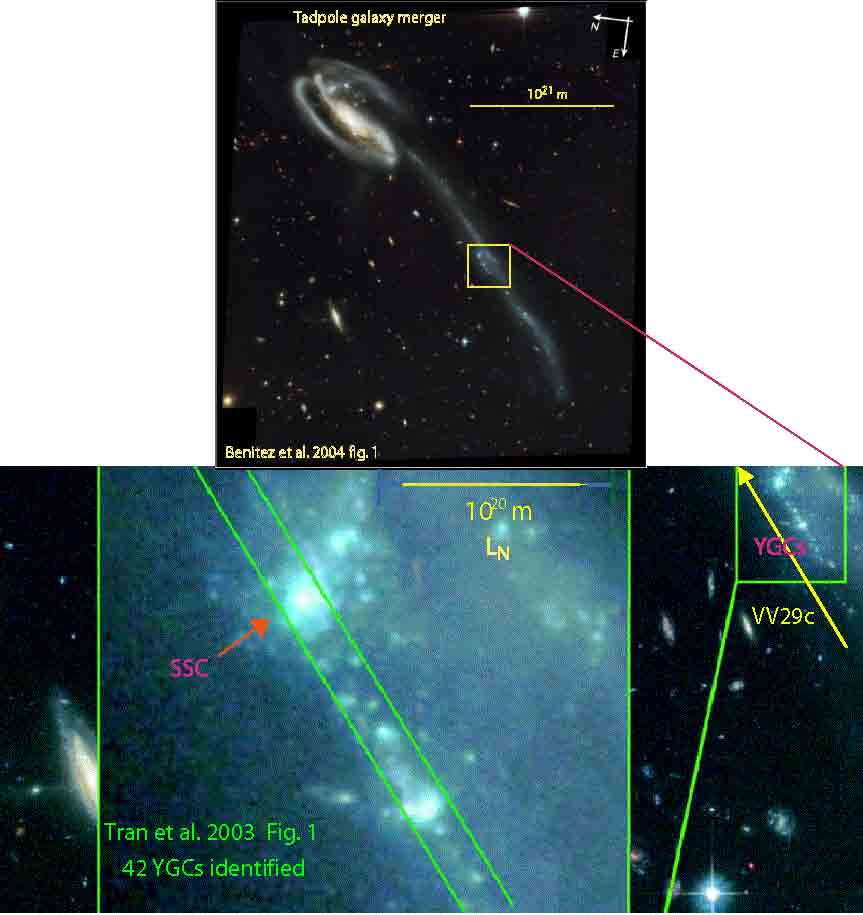}
         \caption{Trail of 42 young-globular-star-clusters (YGCs) 
         in a dark dwarf galaxy examined
spectroscopically by Tran et al. 2003 using the Keck telescope.  The 1 $\arcsec$
Echellette slit and a loose super-star-cluster
(SSC arrow) are shown  at the left.  Ages of the YGCs range
from 3-10 Myr.  The aligned YGC trail is extended by several more YGGs
(arrow on right) and points precisely to the beginning, at $2 \times 10^{21}$ m
distance, of the spiral star wake of VV29c in its capture by VV29a.  The
baryonic dark matter halo of Tadpole is revealed by a looser trail of YGCs
extending to a radius of $4
\times 10^{21}$ m from VV29a, or 130 kpc \citep{gs03a}.}
        \end{figure}

\begin{figure}
         \epsscale{.8}
         \plotone{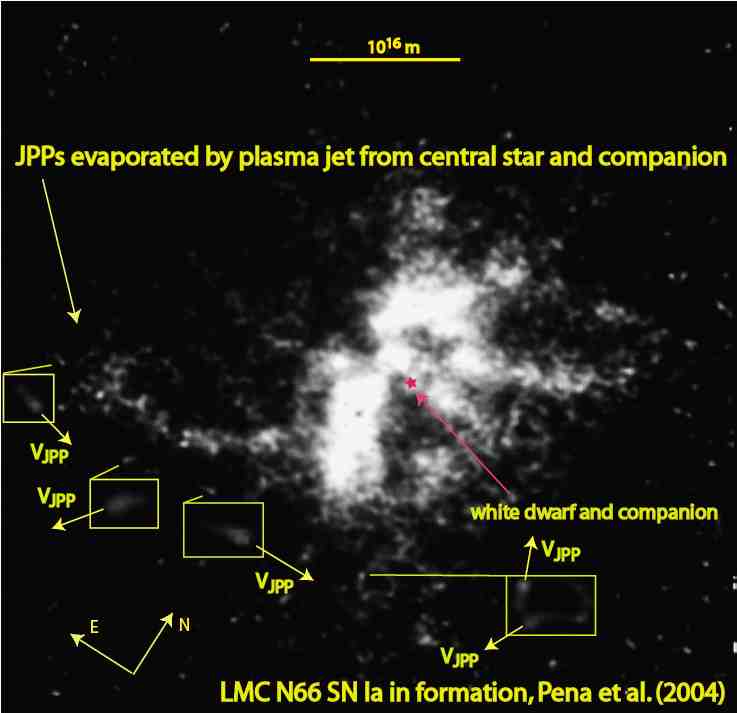}
         \caption{HST/FOC  5007 $\rm \AA$ image (HST archives, 6/26/1991) 
         of distant ($1.7 \times 10^{21}$ m) LMC/N66 PNe with
          a central $1.2 M_{\sun}$ WD-companion close binary rapidly growing toward
           SN Ia formation \citep{pen04}.  Wakes in O III emission (magnified inserts)
            show random  velocities $V_{JPP}$ of the evaporating planets
            in globulette clumps of PFPs \citep{gah07}.  
            From its spectrum the central
             star is a Wolf-Rayet of class WN.  Arc-like patterns show strong
             nutating plasma beams from the binary have
             evaporated $\approx 5 M_{\sun}$ of the ambient JPPs (Fig. 3) .
             From HGD and $\rho_0$ the $2.5 \times 10^{16}$ m radius nebular sphere 
             for LMC/N66 should contain
$\ge 1000 M_{\sun}$ of PFP and JPP Jovian planets.}
        \end{figure}

\begin{figure}
         \epsscale{.8}
         \plotone{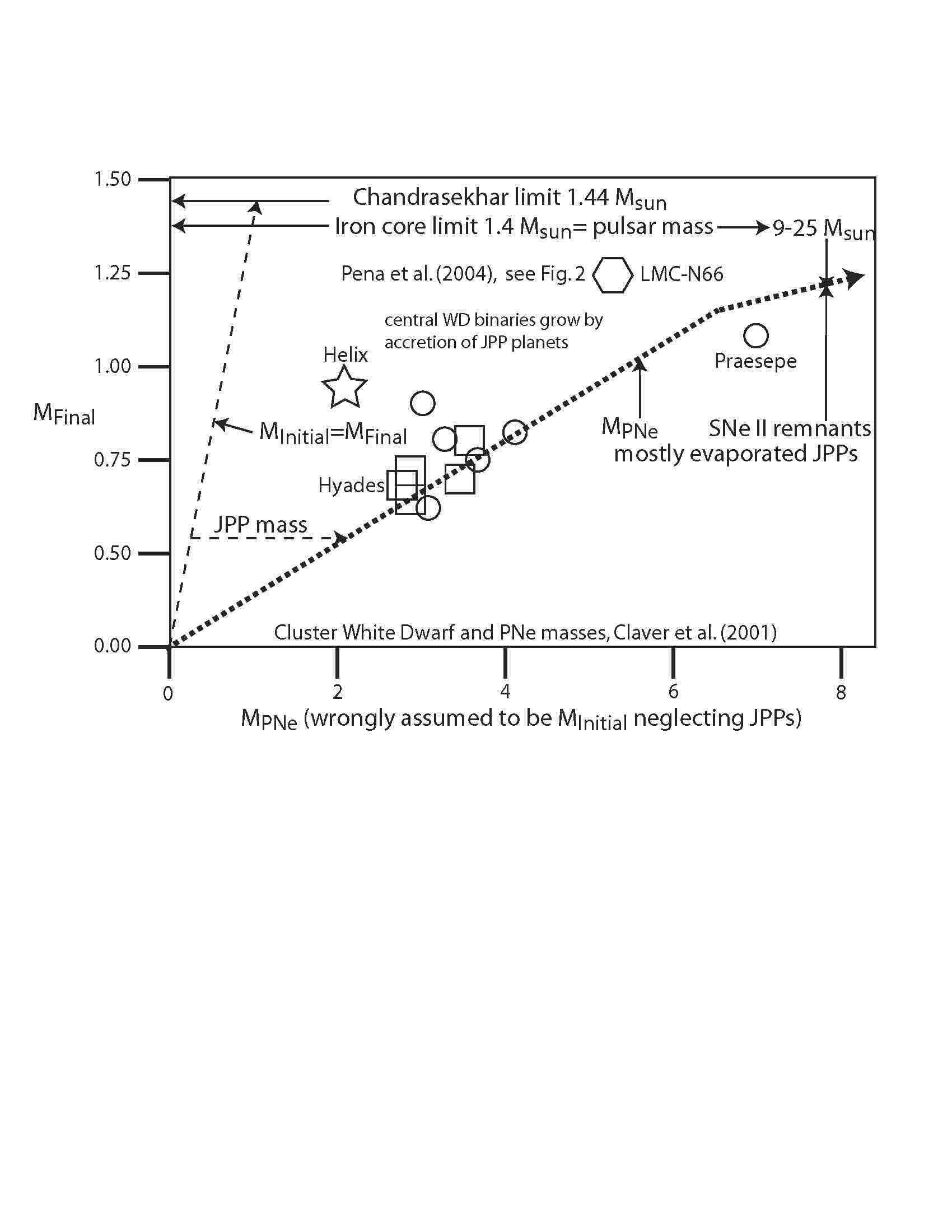}
         \caption{Mass evolution of white dwarf stars to the Chandrasekhar limit by accretion
         of JPPs.  The standard PNe model incorrectly estimates the initial white dwarf mass
          $M_{Initial}$ to be the total PNe mass  $M_{PNe}$, but this includes the brightness
           mass of evaporated
          JPPs.  Measurements for the N66 PNe of Fig. 2 (hexigon)  and the Helix PNe 
          of Figs. 4-6 (star)
        are compared to star cluster PNe of Praesepe (circles) and 
        Hyades (squares) \citep{cla01}.  Infrared detection of JPP atmospheres
        in Helix indicate $M_{PNe} \ge 40 M_{\sun}$ \citep{hor06}. Neutron
        star precursor masses (upper right) should be $\le 1.4 M_{\sun}$ pulsar masses, and much less
        than SN II remnant masses of $(9-25) M_{\sun}$ \citep{gel07,mau04}.}
     
        \end{figure}

\begin{figure}
         \epsscale{0.8}
         \plotone{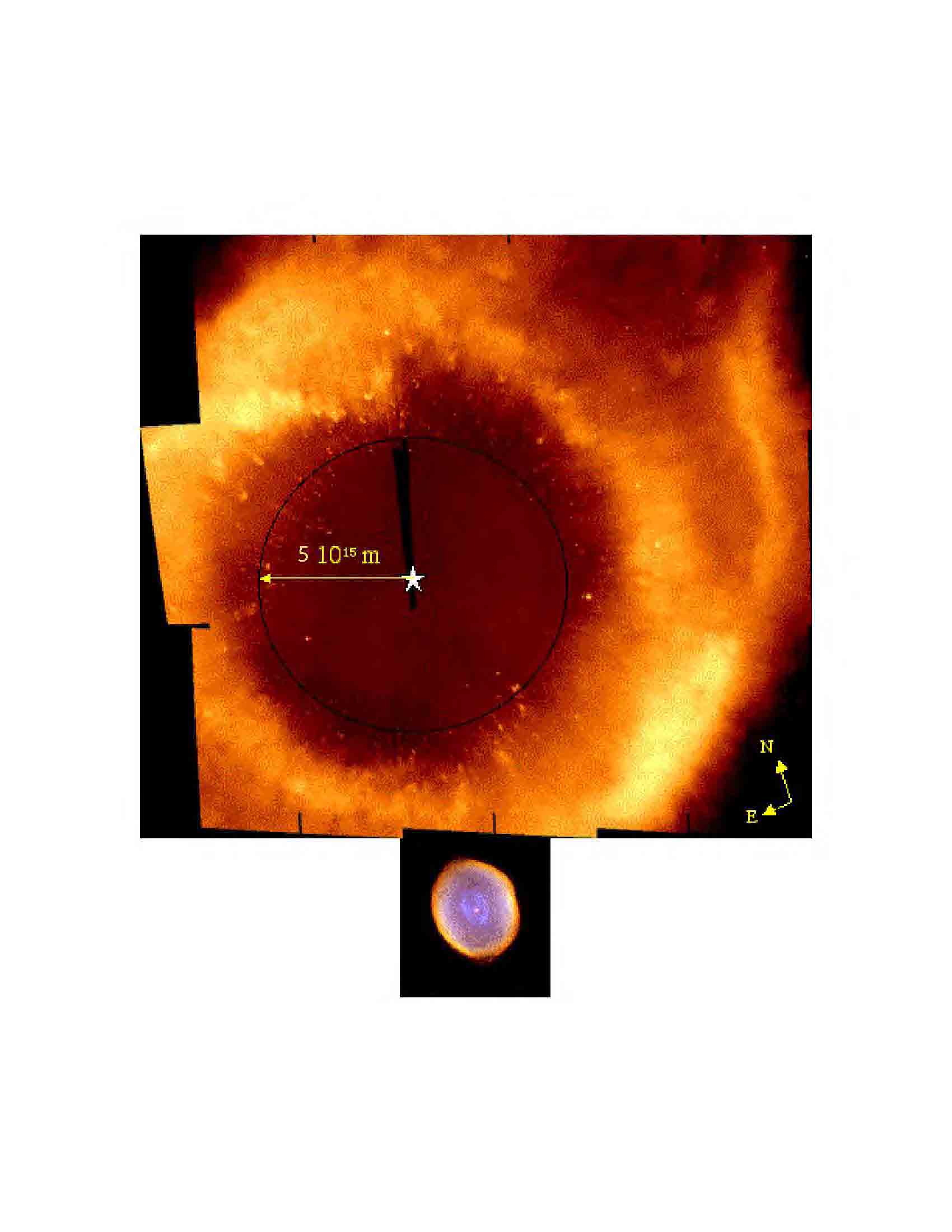}
         \caption{Helix Planetary Nebula HST/ACS/WFC F658N image mosaic.
A sphere with radius $3 \times 10^{15}$ m corresponds to the volume of
primordial-fog-particles (PFPs) with mass density $\rho_0 = 3 \times 10^{-17}$ kg
m$^{-3}$ required to form two central stars by accretion.  The comets
within the sphere are from large gas planets (Jupiters, JPPs) that have
survived evaporation rates of $2 \times 10^{-8} M_{\sun}$/year
\citep{mea98} for the 20,000 year kinematic age of Helix.  The younger planetary
nebula Spirograph (IC 418) shown below with no PFPs is within its accretion
sphere.  From HGD the $2.5 \times 10^{16}$ m radius nebular sphere for Helix contains
$\ge 1000 M_{\sun}$ of dark PFP and JPP planets, from which $ 1.5
\times M_{\sun}$ has been evaporated as detectable  gas and dust
\citep{spe02}.}
        \end{figure}

\begin{figure}
         \epsscale{.8}
         \plotone{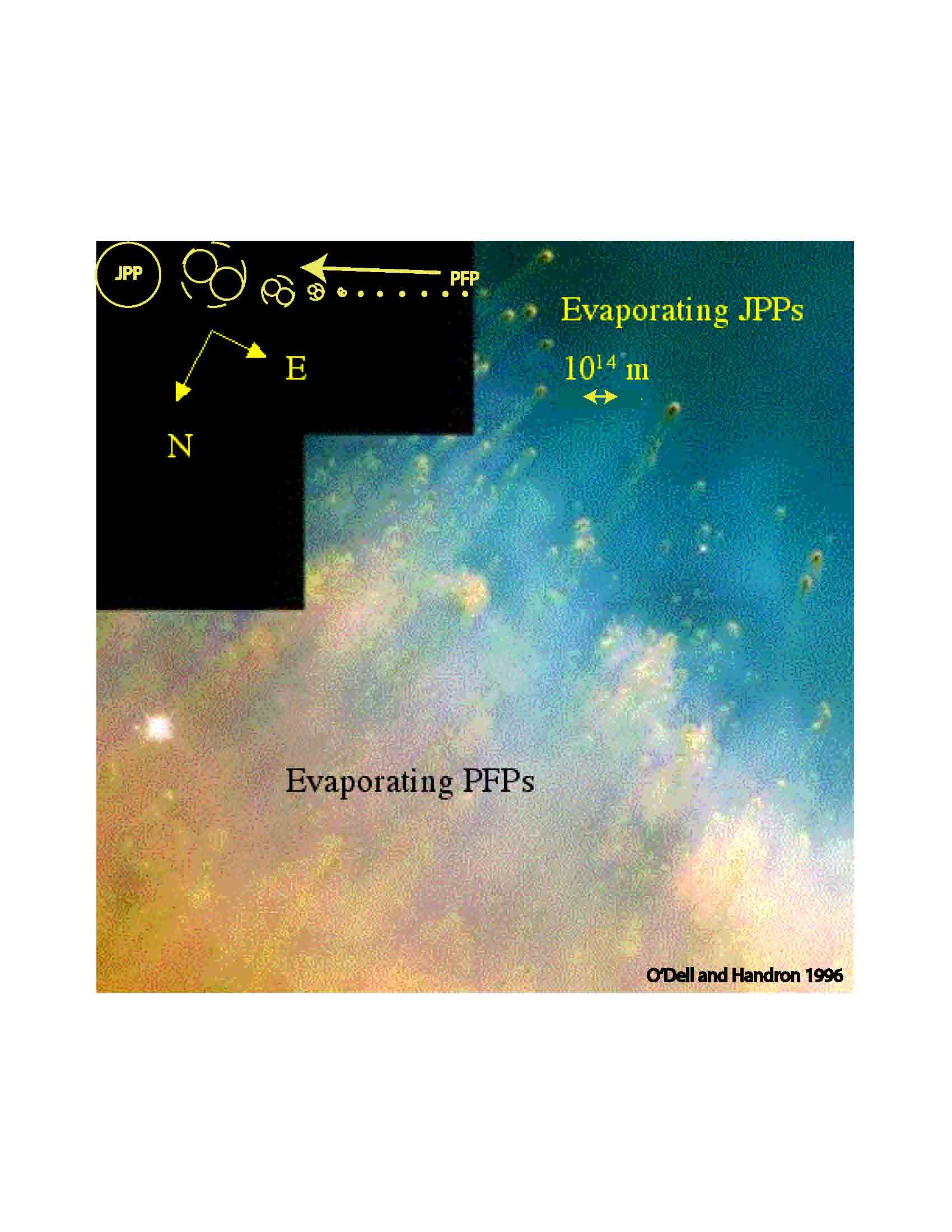}
         \caption{Helix Planetary Nebula HST/WFPC2 1996 image \citep{ode96}
from the strongly illuminated northeast region of Helix containing massive JPP
comets close to the central stars ($\approx 4 \times 10^{15}$ m) with
embedded $\ge 0.4$ Jupiter-mass planet atmospheres.  The largest JPPs
have about Jupiter mass $1.9 \times 10^{27}$ kg from their spacing $\approx 4 \times 10^{14}$ m
assuming primordial density $\rho _0 = 3 \times 10^{-17}$ kg m$^{-3}$.  The $2^{10}$ stage
binary cascade from Earth-mass  to Jupiter-mass is shown in upper left. }
        \end{figure}

\begin{figure}
         \epsscale{.7}
         \plotone{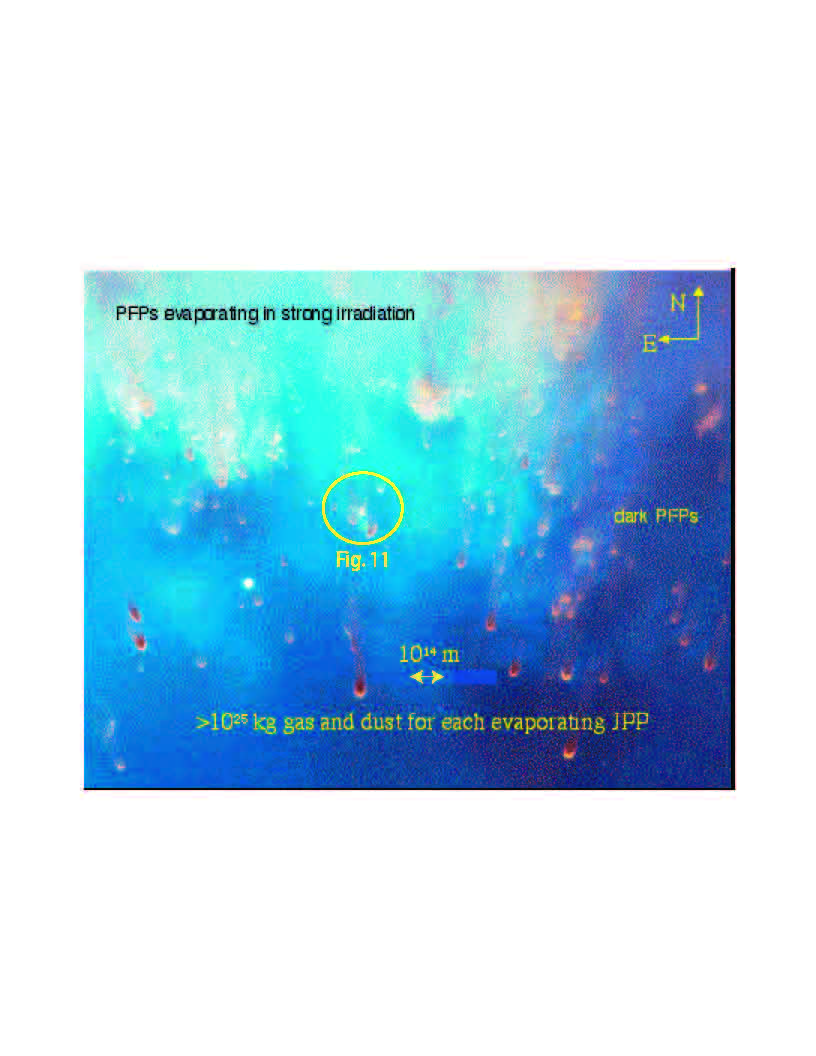}
         \caption{Detail of closely spaced cometary globules to the
north in Helix from the 2002 HST/ACS images at the dark to light transition
marking the clockwise rotation of the beamed radiation from the binary central
star.  Comets (evaporating JPPs) in the dark region to the right have shorter tails and
appear smaller in diameter since they have recently had less intense
radiation than the comets on the left.  Two puffs of gas  deep in
the dark region suggest gravitational collection by planet gravity occured
during the several thousand years since their last time of strong
irradiation. Dark PFPs are detected  in Fig. 11 (circle).}
        \end{figure}

\begin{figure}
         \epsscale{.8}
        \plotone{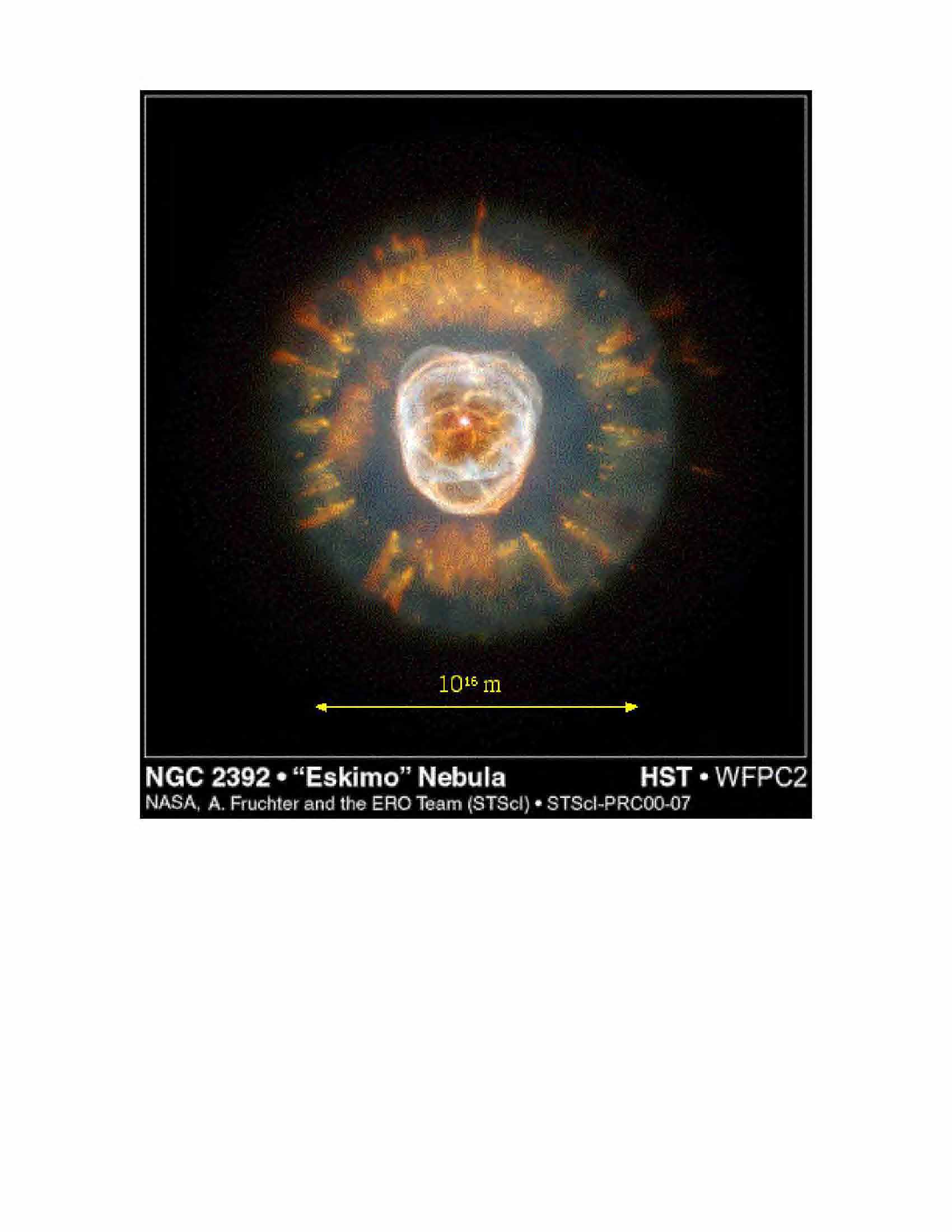}
         \caption{The Eskimo planetary nebula (NGC 2392) is $\approx 12$ times more
distance from earth than Helix, but still shows numerous evaporating PFP and
JPP candidates in its surrounding interstellar medium in the HST/WFPC
images.  The nebula is smaller and younger than Helix, with a central shocked region
like that of Spirograph in Fig. 4.}
        \end{figure}

\begin{figure}
         \epsscale{.7}
        \plotone{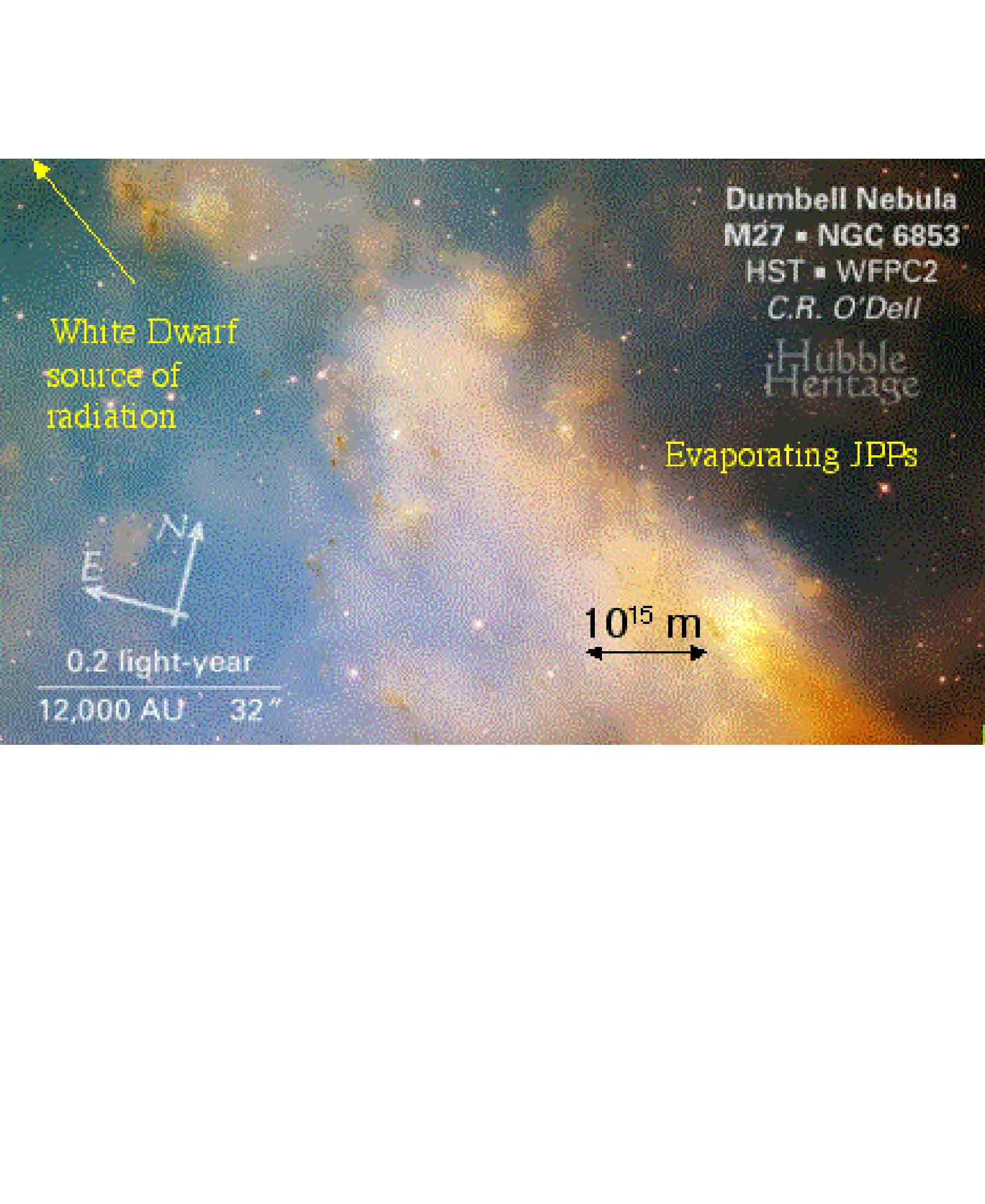}
         \caption{Close-up image of the Dumbbell planetary nebula (M27, NGC
6853) shows numerous closely spaced, evaporating, irradiated PFP and JPP
candidates in its central region.  The PNe is at a distance
$\approx$ 500 pc, with diameter $\approx 2 \times 10^{16}$ m.  The white
dwarf central star
appears to have a companion from the double beamed radiation emitted to
produce the eponymous shape. The lack an apparent accretional hole may be
the result of a different viewing angle (edgewise to the binary star
plane of radiation) than the face-on views of Ring, Helix (Fig. 4) and
Eskimo (Fig. 7).}
        \end{figure}

\begin{figure}
         \epsscale{.6}
         \plotone{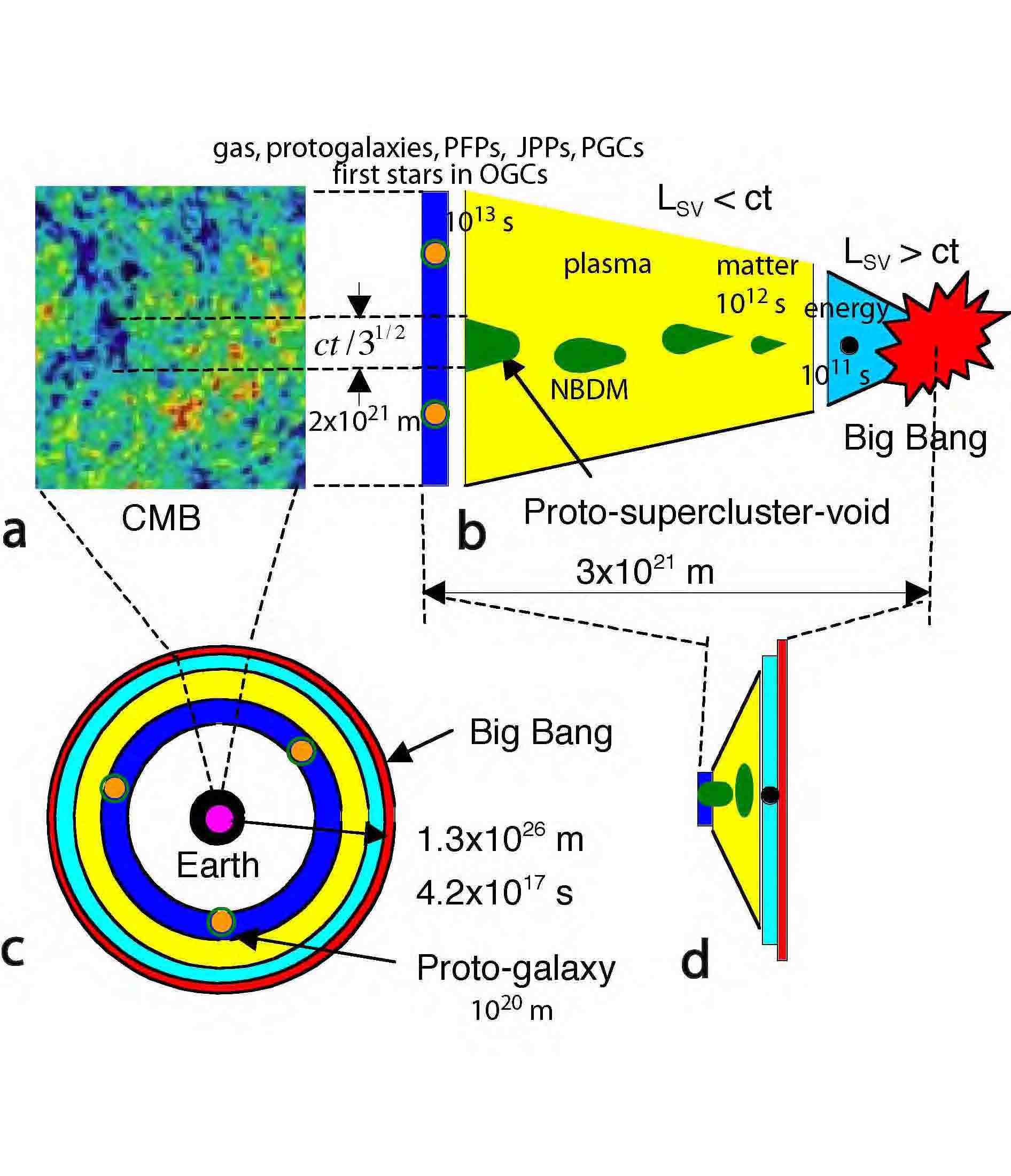}
         \caption{Hydro-Gravitational-Dynamics (HGD) description of the
formation of structure  \citep{gib05, gib04}.  The CMB (a) viewed from the Earth (b) is distant
in both space and time and stretched into a thin spherical shell along
with the energy-plasma  epochs and the big bang (c and d).  Fossils of 
big bang turbulent temperature nucleosynthesize fossil density turbulence patterns in the
H-He density (black dots).  These  trigger gravitational $L_{SV}$ scale 
structures (Tables 1 and 2) in the
plasma epoch as proto-supercluster-voids that  fill with
NBDM (green, probably neutrinos) by diffusion.  The smallest structures emerging
from the plasma epoch are linear chains of fragmented
 $L_N$ scale proto-galaxies.  These fragment into $L_{J}$ scale
PGC clumps of $L_{SV}$ scale PFP Jovian planets that freeze to form the baryonic dark matter
\citep{gib96,sch96} and ``nonlinear grey dust'' sources of SNe Ia
dimming presently misinterpreted as  ``dark energy'' (see Fig. 10).     }
        \end{figure}

\begin{figure}
         \epsscale{.7}
         \plotone{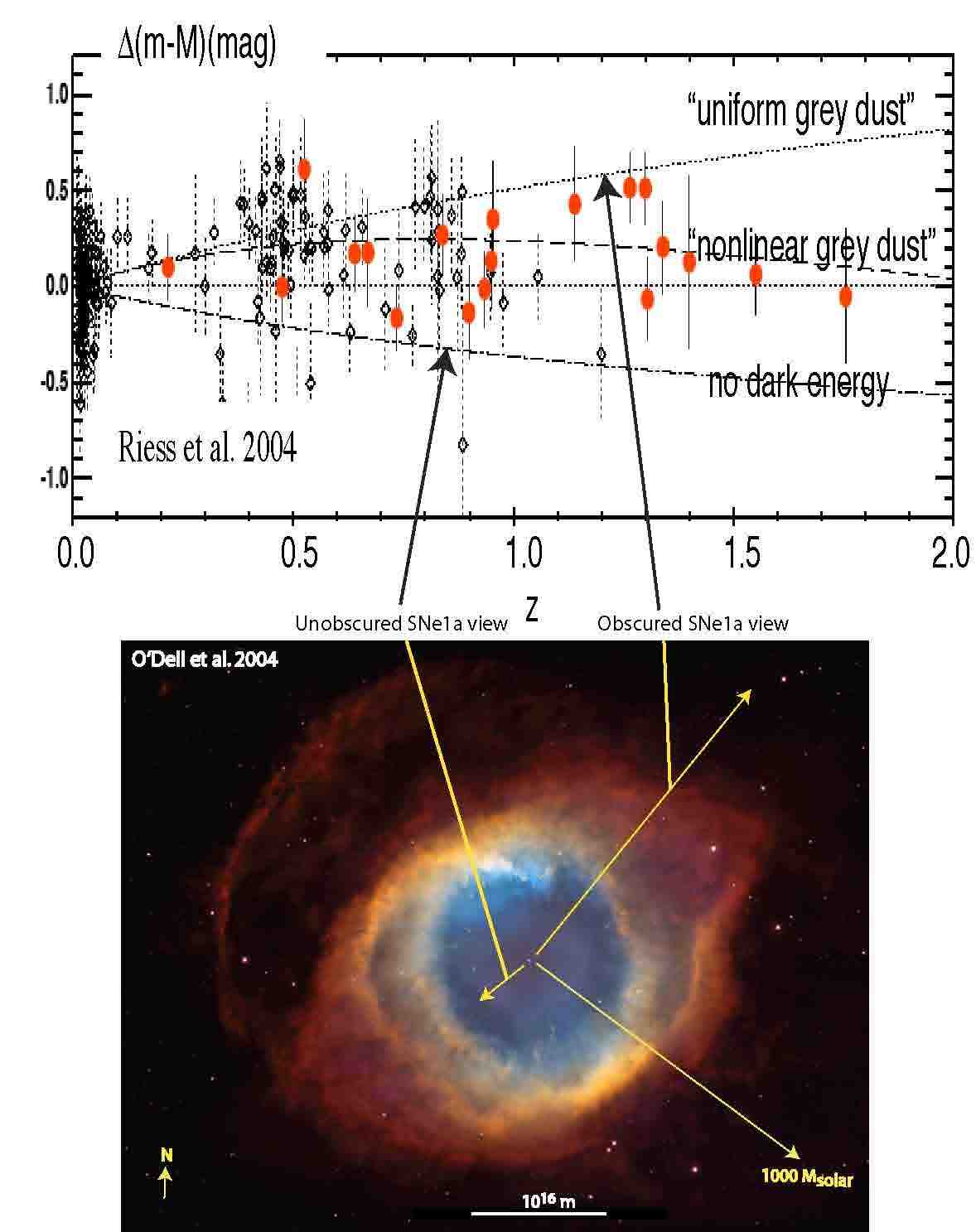}
         \caption{Dimming of SNE Ia magnitudes (top) as a function of
redshift z  \citep{rie04}.  The ``uniform grey dust'' systematic
error is excluded at large z, but the ``nonlinear grey dust''
systematic error from baryonic dark matter PFP and JPP evaporated planet
atmospheres is not, as shown (bottom) Helix PNe \citep{ode04}.  Frozen 
primordial planets comprise the ISM beyond the Oort cavity, which is
 the spherical hole left in a PGC when
the central star is formed by accretion of PFPs and JPP planets.   Planet
atmospheres appear from the dark forming the PNe when the star runs out of fuel and
 the carbon core of the central star contracts.
The rapidly spinning $ 0.7 M_{\sun}$ white dwarf powers axial and equatorial 
plasma jets and winds that  evaporate 
``nonlinear grey dust'' planet atmospheres 
as the white dwarf is fed to $1.44 M_{\sun}$ supernova Ia size by a
 slow rain of planets.  }
        \end{figure}

\begin{figure}
         \epsscale{.7}
        \plotone{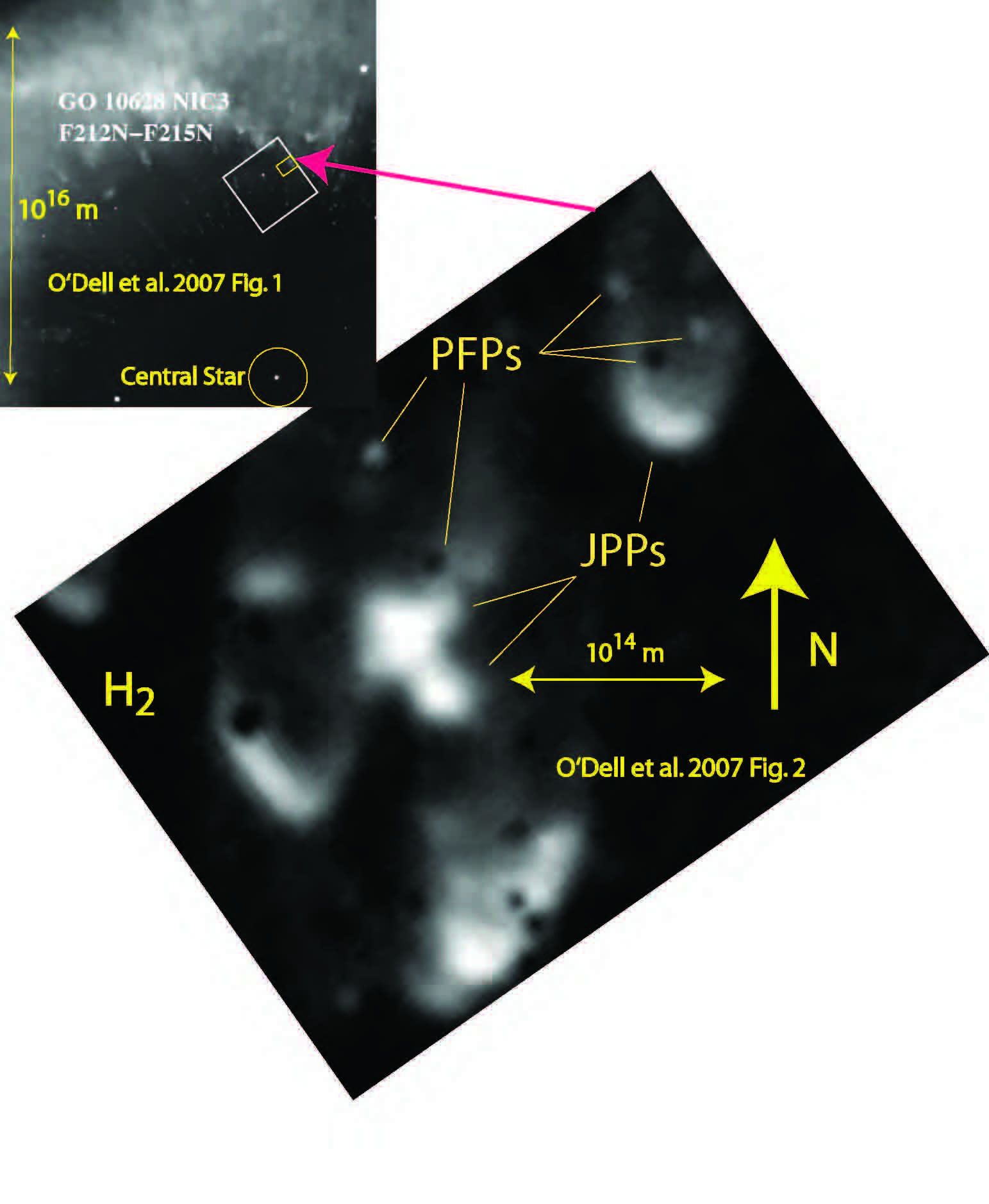}
         \caption{Detail of Helix image in $\rm H_2$ from Figs. 1 and 2 of O'Dell et al. 2007.   Distance
         scales have $\approx 10 \%$ accuracy \citep{har07} by trigonometric parallax,
         giving
         a Helix distance  219 pc with HST pixel  $3.3 \times 10^{12}$ m, the
         size of PFPs identified as the smallest light and dark objects.  
         Such PFPs are not detected closer than $\approx 5 \times 10^{15}$ m from the 
         binary central star (Fig. 4) because they have been evaporated by its radiation and plasma beam.}
        \end{figure}
       
  \begin{figure}
         \epsscale{.7}
         \plotone{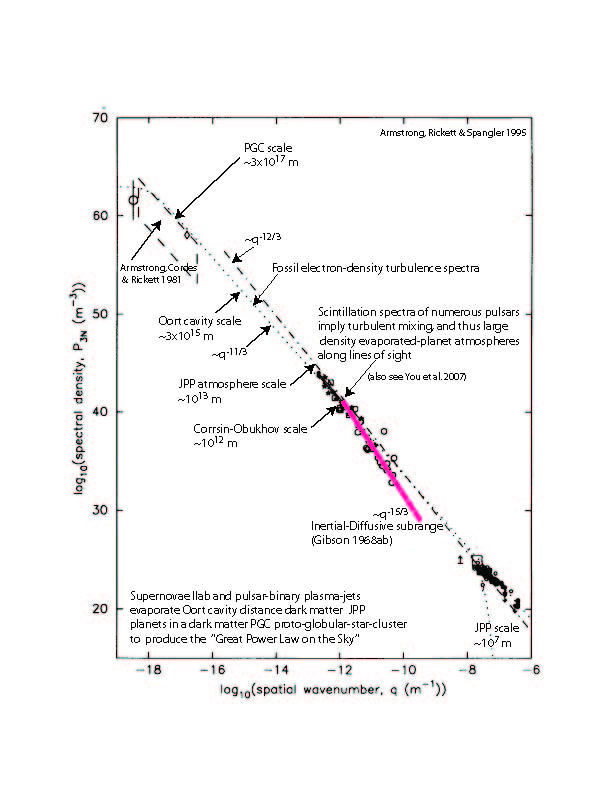}
         \caption{Electron density spectral estimates near Earth
         (within $\approx 10^{19-20}$ m) interpreted using HGD.    The remarkable
         agreement along all lines of sight with the same universal Kolmogorov-Corrsin-Obukov 
         spectral forms reflects the highly uniform behavior of baryonic PCG-PFP dark matter with
          primordial origin in response to SNe II and their pulsars as radio frequency standard candles.  
          From HGD the large stars, supernovae IIab, and neutron-star pulsars of the
          Galaxy disk require pulsar scintillations in dense, previously turbulent,
         JPP planet atmospheres with fossil electron density turbulence.  Dissipation
         rates implied are large, indicating rapid evaporation of the
         frozen gas planets.}
        \end{figure}

\clearpage

\begin{deluxetable}{lrrrrcrrrrr}
\tablewidth{0pt}
\tablecaption{Length scales of self-gravitational structure formation}
\tablehead{
\colhead{Length scale name}& \colhead{Symbol}           &
\colhead{Definition$^a$}      &
\colhead{Physical significance$^b$}           }
\startdata Jeans Acoustic & $L_J$ &$V_S /[\rho G]^{1/2}$& Ideal gas pressure
equilibration

\\Jeans Hydrostatic & $L_{JHS}$ &$[p/\rho^2 G]^{1/2}$& Hydrostatic pressure
equilibration

\\ Schwarz Diffusive & $L_{SD}$&$[D^2 /\rho G]^{1/4}$& $V_D$ balances
$V_{G}$

\\  Schwarz Viscous & $L_{SV}$&$[\gamma \nu /\rho G]^{1/2}$& Viscous force
balances gravitational force

 \\ Schwarz Turbulent & $L_{ST}$&$\varepsilon ^{1/2}/ [\rho G]^{3/4}$&
Turbulence force  balances gravitational force

\\ Kolmogorov Viscous & $L_{K}$&$ [\nu ^3/ \varepsilon]^{1/4}$& Turbulence
force  balances viscous force

\\ Nomura Protogalaxy & $L_N$ & $[L_{ST}]_{CMB}$ & $10^{20}$ m proto-galaxy fragmentation-shape scale

\\ Ozmidov Buoyancy & $L_{R}$&$[\varepsilon/N^3]^{1/2}$& Buoyancy force
balances turbulence force
\\

Gibson Flamelet & $L_{G}$&${v_f } {\gamma} ^{-1}$& Thickness of
flames in turbulent combustion
\\

Particle Collision & $L_{C}$&$ m \sigma ^{-1} \rho ^{-1}$& Distance between
particle collisions
\\

Hubble Horizon & $L_{H}$&$ ct$& Maximum scale of causal connection
\\


\enddata
\tablenotetext{a}{$V_S$ is sound speed, $\rho$ is density, $G$ is Newton's
constant, $D$ is the diffusivity, $V_D \equiv D/L$ is the diffusive velocity
at scale $L$, $V_G \equiv L[\rho G]^{1/2}$ is the gravitational velocity,
$\gamma$ is the strain rate,
$\nu$ is the kinematic viscosity,
$\varepsilon$ is the viscous dissipation rate, $N \equiv
[g\rho^{-1}\partial\rho/\partial z]^{1/2}$ is the stratification frequency,
$g$ is self-gravitational acceleration, $z$ is in the opposite direction
(up),
$v_f$ is the laminar flame velocity,
$m$ is the particle mass,
$\sigma$ is the collision cross section,  $c$ is light speed, $t$ is the
age of universe.}

\tablenotetext{b}{Magnetic and other forces (besides viscous and turbulence)
are negligible for the epoch of primordial self-gravitational structure
formation
\citep{gib96}.}


\end{deluxetable}

\clearpage

\begin{deluxetable}{lrrrrcrrrrr}
\tablewidth{0pt}
\tablecaption{Acronyms}
\tablehead{
\colhead{Acronym}& \colhead{Meaning}           &

\colhead{Physical significance}           }
\startdata

BDM & Baryonic Dark Matter&PGC clumps of JPPs from HGD
\\

CDM & Cold Dark Matter& Questioned concept
\\

CMB & Cosmic Microwave Background&Plasma transition to gas after big bang
\\

HCC & Hierarchical Clustering Cosmology& Questioned CDM concept
\\

HCG & Hickson Compact Galaxy Cluster& Stephan's Quintet (SQ=HGC 92)
\\

HGD & Hydro-Gravitational-Dynamics& Corrects Jeans 1902 theory
\\

ISM &Inter-Stellar Medium& Mostly PFPs and gas from JPPs
\\

JPP&Jovian PFP Planet&H-He planet formed by PFP accretion
\\

$\Lambda$CDMHCC & Dark-Energy CDM HCC&Three questioned concepts
\\

NBDM & Non-Baryonic Dark Matter&Includes (and may be mostly) neutrinos
   \\

OGC & Old Globular star Cluster& PGC that formed stars at $t
\approx 10^{6}$ yr
\\

PFP&Primordial Fog Particle&Earth-mass BDM primordial planet
\\

PGC & Proto-Globular star Cluster& Jeans-mass protogalaxy fragment
\\

SSC & Super-Star Cluster&  A cluster of YGCs
\\

YGC & Young Globular star Cluster& PGC forms stars at $t
\approx$ now
\\


\enddata


\end{deluxetable}

\end{document}